\documentstyle[amssymb,preprint,aps]{revtex}

\begin{document}
\author{M\ J\ Giles}
\address{School of Engineering, University of Northumbria , Ellison Place,Newcastle\\
upon Tyne NE1 8ST, UK}
\date{September 2000}
\title{Anomalous scaling in homogeneous isotropic turbulence}
\maketitle

\begin{abstract}
The anomalous scaling exponents $\zeta _{n}$ of the longitudinal structure
functions $S_{n}$ for homogeneous isotropic turbulence are derived from the
Navier-Stokes equations by using field theoretic methods to develop a low
energy approximation in which the Kolmogorov theory is shown to act
effectively as a mean field theory. \ The corrections to the Kolmogorov
exponents are expressed in terms of the anomalous dimensions of the
composite operators which occur in the definition of $S_{n}$. \ These are
calculated from the anomalous scaling of the appropriate class of nonlinear
Green's function, using an $uv$ fixed point of the renormalisation group,
which thereby establishes the connection with the dynamics of the
turbulence. \ The main result is an algebraic expression for $\zeta _{n}$,
which contains no adjustable constants. \ It is valid at orders $n$ below $%
g_{\ast }^{-1}$, where $g_{\ast }$ is the fixed point coupling constant. \
This expression is used to calculate $\zeta _{n}$ for orders in the range $%
n=2$ to $10$, and the results are shown to be in good agreement with
experimental data, key examples being $\zeta _{2}=0.7$, $\zeta _{3}=1$ and $%
\zeta _{6}=1.8$.
\end{abstract}

\section{Introduction}

\label{section1}

The study of homogeneous isotropic turbulence has as its aim the derivation
of the statistical features of small scale velocity fluctuations at high
Reynolds numbers, based on the assumption that they exhibit universal
characteristics independent of the form of the large scale flow structures
[1-3]. A key quantity of interest is the longitudinal velocity increment, $%
v_{+}-v_{-},$ where $v_{\pm }=v_{1}(x\pm r/2,y,z,t),$ the velocity component 
$v_{1}$ and the separation distance $r$ both being in the same direction,
here the $x$-axis. An empirical fact is that its $n$th order moment, the
longitudinal structure function $S_{n}(r)$, defined by 
\begin{equation}
S_{n}(r)=\left\langle (v_{+}-v_{-})^{n}\right\rangle ,  \label{a1}
\end{equation}
exhibits multiscaling. That is, the exponent $\zeta _{n}$, defined by the
scaling relation 
\begin{equation}
S_{n}(r)\sim r^{\zeta _{n}},  \label{a2}
\end{equation}
is a nonlinear function of the order $n.$ This behaviour is not explained by
the classical turbulence theory of Kolmogorov [4] which yields a linear
dependence 
\begin{equation}
\zeta _{n}^{Kol}=\frac{n}{3}.  \label{a3}
\end{equation}
Moreover, the amount by which $\zeta _{n}$ differs from $\zeta _{n}^{Kol},$
called the anomaly, has proved stubbornly resistant to attempts at
quantitative explanation [1-3,5,6]. The obstacle to progress with the theory
is the strong nonlinearity of the governing Navier-Stokes (NS) equations. In
this paper, our aim is to show how modern statistical field theory can be
used to overcome this difficulty and provide theoretical predictions for $%
\zeta _{n}$, which agree well with turbulent flow data.

The idea that statistical field theory can be brought to bear on the problem
of turbulence is not itself new. Indeed, interest in describing turbulence
in terms of the underlying functional probability distribution of the
velocity field, together with its corresponding generating functional $W,$
has a long history [5,6]. But such work has suffered from the weakness of
relying on conventional perturbation theory to effect closure of the
statistical hierarchy, whereas it is widely believed that a non-perturbative
treatment is necessary, because the NS equations lack a small parameter.\
Consequently, progress with this approach has been disappointing.

The question is whether we can find a middle course, which avoids the
limitations of conventional perturbation theory, while not demanding an
intractable non-perturbative approach. Here we explore the possibility of
formulating a more efficient perturbation theory by developing a zero-order
solution which already accounts for the dominant nonlinear interactions, in
an attempt, as it were, to deplete the effect of the nonlinearity. We shall
do this by adopting a more general quadratic form in $W$ in place of the
viscous form which arises naturally. The modified quadratic form is
determined self-consistently from the NS nonlinearity using the linear
response function and the energy equation. In the inertial range, it leads
directly to the Kolmogorov distribution, after allowing for the kinematic
effect of the sweeping of the smaller scales by the larger ones. The
difference between these quadratic forms then appears as a perturbation,
which, as we shall see, is not critical, provided that the force spectrum
function is non-zero only at small wavenumbers and yields a finite input
power.

Having incorporated the dominant nonlinearities which are responsible for
the turbulence energy cascade into the zero-order solution, what is then
lacking is the effect of the fluctuating dissipation rate, which is the
well-known defect of the Kolmogorov theory [5]. In this approach, the
perturbation theory is then, in effect, only required to accommodate the
residual coupling associated with these fluctuations, which are directly
responsible for the anomalies. The fact that the anomalies are small, and
associated with a weak residual coupling, provides good reason to expect
that a small expansion parameter might emerge, thereby rendering the problem
accessible to perturbation theory, essentially by means of a standard
loop-expansion of the generating functional.

Although the use of the modified quadratic form as an initial approximation
would appear to be an attractive option, providing a sound physical basis
for the approximate evaluation of the generating functional, it poses severe
technical problems, the most significant being the occurrence of divergences
at higher orders in perturbation theory, due to the incomplete
representation of the large scale flow. These divergences are of two types:
power divergences (including power $\times $ logarithmic), which are
associated with the sweeping and pure logarithmic divergences, which
describe the cascade process.\ On the other hand, statistical field theory
[7], provides the mathematical techniques needed to compensate for such
divergences, in the form of the well-known processes of resummation and
renormalisation. In particular, renormalisation [8] provides a procedure
whereby the scale invariance in (\ref{a2}) can be recovered from a divergent
theory, yielding the exponents in terms of the anomalous dimensions of the
composite operators appearing in (\ref{a1}), which we can calculate from the
appropriate nonlinear Green's functions. The modified quadratic form itself
follows uniquely from the requirement for the absence of non-renormalisable
terms, after renormalising the basic parameters of $W$ and allowing for
sweeping.

Fortunately, sweeping effects do not pose an insuperable obstacle,
notwithstanding that the initial formulation is Eulerian. Indeed, we show
that the power divergences associated with sweeping can be removed by
introducing a single sweeping interaction term, which can be derived from
the generating functional itself using a random Galilean transformation of
the velocity field, having an {\it rms} convection speed which is calculated
from the NS nonlinearity. The application of this transformation does not,
of course, affect the values of $S_{n}(r)$ and, thus, enables the straining
interactions which determine the spectrum to be separated from the
background of sweeping convection, yielding, in effect, quasi-Lagrangian
forms. \ In this way, as we shall show below, it proves possible to
demonstrate multiscaling and calculate the anomalies of the structure
functions accurately.

\section{Theoretical Foundations}

\label{section2}

Our starting point is the NS equations, describing flow in an incompressible
fluid of unit density, velocity ${\bf v},$ kinematic viscosity $\overline{%
\nu }$ and pressure $p$, and driven by a random solenoidal stirring force $%
{\bf f}$, which are 
\begin{equation}
\frac{\partial {\bf v}}{\partial t}+{\bf v\cdot }\nabla {\bf v}=-\nabla p+%
\bar{\nu}\nabla ^{2}{\bf v}+{\bf f},  \label{a4}
\end{equation}
and

\begin{equation}
\mathop{\rm div}%
{\bf v}=0.  \label{a5}
\end{equation}

Suppose that

\begin{equation}
{\bf v}(\hat{x})={\bf V}(\hat{x}|{\bf f})  \label{a6}
\end{equation}
is the solution of (\ref{a4}) at the space-time point $\hat{x}=({\bf x},t),$
corresponding to a force ${\bf f}(\hat{x})$, which has a Gaussian
probability distribution ${\cal P}({\bf f}).$ Then the generating functional 
$W$ for the correlation functions of the velocity field can be written as
the functional integral 
\begin{equation}
W=\int \exp (S){\cal P}({\bf f})\,{\cal D}{\bf f},  \label{a7}
\end{equation}
where the source term is given by 
\begin{equation}
S({\bf J})=%
\textstyle\int%
{\bf J}(\hat{x})\cdot {\bf V}(\hat{x}|{\bf f})\,d\hat{x},  \label{a8}
\end{equation}
and the correlators follow by functional differentiation with respect to the
source field ${\bf J(}\hat{x}).$ Given that we cannot obtain an explicit
expression for the solution (\ref{a6}), the crux of the problem is how to
approximate (\ref{a7}) with the accuracy required to calculate the $\zeta
_{n}$. In our approach, as indicated above, we prove that the Kolmogorov
theory can be used effectively as a mean field theory in a saddle-point
evaluation of (\ref{a7}), and that this leads to an expansion which has a
genuinely small coupling constant. \ 

Within the context of a field theoretic interpretation of (\ref{a7}), each
term of the binomial expansion of (\ref{a1}) must be regarded as an operator
product of the usual Wilson type, (see eg [7,8] ). Correspondingly, the
powers of $v_{\pm }$ must be treated as composite operators, which, in
accordance with standard procedures [7], must be generated from $W$ by
independent sources. Here, our aim is to limit the composite operators that
need to be allowed for to those which appear explicitly in the definition of 
$S_{n}(r),$ as given in (\ref{a1}). To this end, we define a set of
longitudinal composite operators $O_{s}(\hat{x}),$ for $s=2,3,4,...,$ by 
\begin{equation}
O_{s}(\hat{x})=v_{1}(\hat{x})^{s}/s!,  \label{a9}
\end{equation}
which we generate from $W$ by adding to the source term (\ref{a8}), the
additional term 
\begin{equation}
-\sum_{s}\int t_{s}(\hat{x})O_{s}(\hat{x})d\hat{x}.  \label{a10}
\end{equation}

We also need to include in the definition of $W$ a means of establishing the
vital link between the time-independent definition of $S_{n}(r)$ and the
dynamics of the turbulence. This requires the introduction of a dynamic
response operator, which we define to be the functional differentiation
operator 
\begin{equation}
{\cal F}_{\alpha }(\hat{x})=\frac{\delta }{i\delta f_{\alpha }(\hat{x})}.
\label{a11}
\end{equation}
Its inclusion in the definition of $W$ adds a final source term to $S,$
given by 
\begin{equation}
\int J_{\alpha }(\hat{x}){\cal F}_{\alpha }(\hat{x})d\hat{x},  \label{a12}
\end{equation}
where summation over repeated vector indices is implied here and below.

The terms (\ref{a8}), (\ref{a10}) and (\ref{a12}) together constitute the
full source term for (\ref{a7}) which becomes, therefore, 
\begin{equation}
S({\bf J,}\widetilde{{\bf J}},t_{s})=\int \{J_{\alpha }(\hat{x})V_{\alpha }(%
\hat{x})+\widetilde{J}_{\alpha }(\hat{x}){\cal F}_{\alpha }(\hat{x}%
)-\sum_{s}t_{s}(\hat{x})O_{s}(\hat{x})\}d\hat{x},  \label{a13}
\end{equation}
and this completes the definition of $W.$ Thus, (\ref{a7}) and (\ref{a13})
provide the foundation of our approach to the calculation of $\zeta _{n}$.
However, before we proceed with this calculation, we need to cast $W$ into a
conventional field theory form, and introduce the modified quadratic form.

A straightforward method of transforming (\ref{a7}) into a conventional
field theory form is to replace ${\cal P}({\bf f})$ by its functional
Fourier transform and then integrate over ${\bf f}$. This is the stage at
which we make explicit use of the NS equations. Essentially, to effect the
transformation, we change our perspective by replacing the velocity field $%
{\bf V(f)}$ generated by the force ${\bf f,}$by the force ${\bf F(v)}$
needed to excite a particular realisation ${\bf v}$ of the flow field. The
operator (\ref{a11}) is then replaced by an equivalent conjugate vector
field $\widetilde{{\bf v}}$.

To carry out this transformation, we work in the Fourier domain setting 
\[
{\bf v}(\hat{x})=\int \exp (i\hat{k}\cdot \hat{x})\,{\bf v}(\hat{k})D\hat{k}%
, 
\]
where $\hat{k}$ denotes $({\bf k,}\omega )$, so that $\hat{k}\cdot \hat{x}%
=\omega t-{\bf k}\cdot {\bf x}$, while $D\hat{k}=d\omega d{\bf k}/(2\pi
)^{4}.$ Then, from (\ref{a4}) and (\ref{a5}), we have 
\begin{equation}
F_{\alpha }(\hat{k},{\bf v})=G_{0}(\hat{k})^{-1}v_{\alpha }(\hat{k})-\frac{i%
}{2}(2\pi )^{4}P_{\alpha \beta \gamma }({\bf k})\int v_{\beta }(\hat{p}%
)v_{\gamma }(\hat{q})\delta (\hat{p}+\hat{q}-\hat{k})D\hat{p}D\hat{q}.
\label{a15}
\end{equation}
The notation here is the following. $G_{0}(\hat{k})$ is the zero-order
approximation to the response function $G(\hat{k})$ defined below in (\ref
{a87}) and (\ref{a92}), namely 
\begin{equation}
G_{0}(\hat{k})=\frac{1}{i\omega +\tau _{\nu }(k)^{-1}},  \label{a16}
\end{equation}
where 
\[
\tau _{\nu }(k)^{-1}=\bar{\nu}k^{2}. 
\]
$P_{\alpha \beta \gamma }({\bf k})$ is the NS vertex defined by 
\[
P_{\alpha \beta \gamma }({\bf k})=k_{\beta }P_{\alpha \gamma }({\bf k}%
)+k_{\gamma }P_{\alpha \beta }({\bf k}). 
\]
where 
\[
P_{\alpha \beta }({\bf k})=\delta _{\alpha \beta }-k_{\alpha }k_{\beta
}/k^{2}. 
\]

Next we write the Gaussian distribution of{\bf \ }${\bf f}${\bf \ }in the
form 
\begin{equation}
{\cal P}({\bf f})={\cal N}\exp \left\{ -\frac{1}{2}\int f_{\alpha }(-\hat{k}%
)h(k)^{-1}P_{\alpha \beta }({\bf k})f_{\beta }(\hat{k})D\hat{k}\right\} ,
\label{a20}
\end{equation}
for which the corresponding force covariance is 
\[
\left\langle f_{\alpha }(\hat{k})f_{\beta }(\hat{l})\right\rangle =(2\pi
)^{4}\delta (\hat{k}+\hat{l})h(k)P_{\alpha \beta }({\bf k}), 
\]
where the force spectrum function $h(k)$ is an arbitrary function which is
assumed to be peaked near the origin so that the power input $\int h(k)d{\bf %
k}$ is finite.\ We now change the functional integration over ${\bf f}$ in (%
\ref{a7}) to an integration over ${\bf v}$ by means of the transformation $%
{\bf v}(\hat{k})={\bf V}(\hat{k}|{\bf f})$, and substitute the
representation 
\begin{equation}
{\cal P}({\bf f})={\cal N}\int \exp \left\{ -\frac{1}{2}\int \tilde{v}%
_{\alpha }(-\hat{k})h(k)P_{\alpha \beta }({\bf k)}\tilde{v}_{\beta }(\hat{k}%
)D\hat{k}+i\int \tilde{v}_{\alpha }(-\hat{k})f_{\alpha }(\hat{k})D\hat{k}%
\right\} {\cal D}\widetilde{{\bf v}},  \label{a22}
\end{equation}
Since the Jacobian only contributes an unimportant constant, we get 
\begin{equation}
W({\bf J},\widetilde{{\bf J}},t_{s})=\int \exp \left[ -L({\bf v},\widetilde{%
{\bf v}})+S({\bf J},\widetilde{{\bf J}},t_{s})\right] {\cal D}{\bf v}\,{\cal %
D}\widetilde{{\bf v}},  \label{a23}
\end{equation}
where 
\begin{equation}
L({\bf v},\widetilde{{\bf v}})=\frac{1}{2}\int \tilde{v}_{\alpha }(-\hat{k}%
)h(k)P_{\alpha \beta }({\bf k})\tilde{v}_{\alpha }(\hat{k})D\hat{k}-i\int 
\tilde{v}_{\alpha }(-\hat{k})F_{\alpha }(\hat{k},{\bf v})D\hat{k},
\label{a24}
\end{equation}
while the source term (\ref{a13}) becomes 
\begin{equation}
S({\bf J,}\widetilde{{\bf J}},t_{s})=\int \{J_{\alpha }(-\hat{k})v_{\alpha }(%
\hat{k})+\tilde{J}_{\alpha }(-\hat{k})\tilde{v}_{\alpha }(\hat{k}%
)-\sum_{s}t_{s}(-\hat{k})O_{s}(\hat{k})\}D\hat{k},  \label{a25}
\end{equation}
The expression (\ref{a23}) casts $W$ into the form of an MSR type functional
integral [9].

Now the quadratic form appearing in (\ref{a24}) does not provide a good
initial approximation for inertial range scaling because, of course, it
merely describes the viscous decay of an externally driven random flow, with
no account taken of the nonlinear interactions. It is thus essential in
developing an expansion theorem for (\ref{a23}) to introduce a more
appropriate quadratic form. Now the general theory of quadratic forms in a
Hilbert space indicates that we can introduce at most two functions.These
can be taken as an apparent force spectrum $D_{0}(k)$ and an effective micro
timescale $\tau _{0}(k)$, which are related to the energy in wavemode ${\bf k%
}$, $Q(k)$, by 
\begin{equation}
Q(k)=\tau _{0}(k)D_{0}(k).  \label{a180}
\end{equation}
The modified quadratic form in $L({\bf v,\tilde{v})}$ is then obtained,
firstly, by replacing $h(k)$ with $D_{0}(k)$ and, secondly, by replacing the
viscous timescale $\tau _{\nu }(k)$ by the the effective timescale $\tau
_{0}(k)$, so that the viscous propagator (\ref{a16}) in (\ref{a15}) is
replaced by 
\[
G_{0}(k)=\frac{1}{i\omega +\tau _{0}(k)^{-1}}. 
\]
Thus, we now have in place of (\ref{a24}) 
\begin{equation}
L({\bf v},\widetilde{{\bf v}})=\frac{1}{2}\int \tilde{v}_{\alpha }(-\hat{k}%
)D_{0}(k)P_{\alpha \beta }({\bf k})\tilde{v}_{\beta }(\hat{k})D\hat{k}-i\int 
\tilde{v}_{\alpha }(-\hat{k})F_{\alpha }(\hat{k},{\bf v})D\hat{k},
\label{a181}
\end{equation}
in which $D_{0}(k)$ and $\tau _{0}(k)$ are, as yet, unknown functions to be
determined in an appropriate way from the energy equation and the linear
response function. The idea that one should replace the viscous quadratic
form by a modified form was suggested originally in [10], where it was used
in conjunction with a variational principle based on an entropy
functional,but recent work [11] has shown that this approach contains an
arbitrary element. However, we shall not need to invoke any additional
principle, because we shall be able to deduce the modified quadratic form in
a self-consistent way from the 1-loop expansion, as we have already
indicated.

The introduction of the modified quadratic form as a basis for an expansion
theorem for (\ref{a23}) requires the inclusion of the difference terms as
perturbations, which contributes an additional term to $L$ given by 
\begin{equation}
\Delta L_{0}=\frac{1}{2}\int \tilde{v}_{\alpha }(-\hat{k})\{h(k)-D_{0}(k)%
\}P_{\alpha \beta }({\bf k})\tilde{v}_{\beta }(\hat{k})D\hat{k}-i\int \tilde{%
v}_{\alpha }(-\hat{k})\{\tau _{\nu }(k)^{-1}-\tau _{0}(k)^{-1}\}v_{\alpha }(%
\hat{k})D\hat{k}.  \label{a182}
\end{equation}
These terms have the same form as the counterterms introduced below in (\ref
{a183}) to accommodate the pure logarithmic divergences but their role, as
we shall see, is not critical as regards calculating the inertial range
exponents.

The derivation of the functions $D_{0}(k)$ and $\tau _{0}(k)$ occurring in
the modified quadratic form entails a detailed discussion of sweeping
convection, the structure of the Feynman diagrams associated with the loop
expansion of $W$ and the establishment of the condition for the absence from
the linear response function of non-renormalisable terms. We shall defer
detailed discussion of these topics until Sections \ref{section7} and \ref
{section8} and, meanwhile, proceed with the calculation of the anomalous
exponents by anticipating their forms, which, in the inertial range, are 
\begin{equation}
D_{0}(k)=D_{0}k^{-3},  \label{a184}
\end{equation}
and 
\begin{equation}
\tau _{0}(k)^{-1}=\nu _{0}k^{2/3}.  \label{a185}
\end{equation}
Clearly, these forms imply that the zero order solution behaves in the
inertial range {\it as if} the fluid were stirred by a random force with a $%
k^{-3}$ force correlation spectrum and responds to it with a Lagrangian time
scale $\varpropto k^{-2/3}$. Thus, they lead to the Kolmogorov
distribution.We shall explain how this result follows from the generating
functional in Section \ref{section8}. The advantage of this approximation is
that it achieves a prime requirement of any efficient perturbation theory,
which is a zero-order approximation that already closely approximates the
desired solution. \ 

On the other hand, as we have indicated, the resulting perturbation theory
yields divergences at higher orders. But these divergences can be handled by
standard renormalisation procedures. Fortunately, as regards the calculation
of $\zeta _{n}$, we need consider only logarithmic divergences. As discussed
above, this is because the power divergences represent the kinematic effect
of the sweeping of small scales by larger scales. Indeed, as we shall show
in Section \ref{section7}, such terms are precisely those which can be
generated by applying a random Galilean transformation of the velocity field
to $W.$ Consequently, they can be cancelled by introducing the appropriate
vertex into $W$, yielding quasi-Lagrangian approximations. Hence, from a
purely practical calculational point of view, the effect of sweeping can be
removed from the calculation of $\zeta _{n}$ simply by discarding power
divergences. We are then left with the logarithmic divergences, which we can
sum by renormalisation group methods.

Thus, an important implication of using the modified quadratic form as an
initial approximation for the calculation of $\zeta _{n}$ is that
renormalisation becomes a necessary preliminary. So we need to identify the
counterterms which arise in $W$ under renormalisation and obtain the
transformation rule which connects the bare and renormalised generating
functionals. Renormalisation is applied to the viscosity and force constants
appearing in (\ref{a184}) and (\ref{a185}) in the usual way by introducing
renormalisation constants $Z_{\nu }$ and $Z_{D}$, which relate their bare
values $\nu _{0}$ and $D_{0}$ to their renormalised replacements $\nu $ and $%
D$ by 
\begin{equation}
\nu _{0}=\nu Z_{\nu }\text{ \ and\ \ }D_{0}=DZ_{D}.  \label{a28}
\end{equation}
This generates counterterms in (\ref{a24}) for the elementary fields $({\bf v%
}$ and $\widetilde{{\bf v}})$ given by 
\begin{eqnarray}
\Delta L_{ef} &=&-\Delta Z_{\nu }i\int \tilde{v}_{\alpha }(-\hat{k})\tau
(k)^{-1}v_{\alpha }(\hat{k})D\hat{k}  \label{a183} \\
&&+\Delta Z_{D}\frac{1}{2}\int \tilde{v}_{\alpha }(-\hat{k})D(k)P_{\alpha
\beta }({\bf k})\tilde{v}_{\alpha }(\hat{k})D\hat{k},  \nonumber
\end{eqnarray}
where we have defined renormalisation constant increments by 
\[
\Delta Z_{\nu ,D}=Z_{\nu ,D}-1. 
\]
The additional renormalisation which must be applied to the composite
operators (\ref{a9}) also takes the standard form 
\begin{equation}
(O_{s})_{B}=Z_{s}(O_{s})_{R}.  \label{a31}
\end{equation}
The corresponding counterterm is obtained by substituting (\ref{a31}) in (%
\ref{a25}) to get 
\[
\Delta L_{co}=\sum_{s}\Delta Z_{s}\int t_{s}(-\hat{k})O_{s}(\hat{k})D\hat{k}%
, 
\]
where 
\[
\Delta Z_{s}=Z_{s}-1. 
\]

We conclude this section by giving the transformation which relates the
generating functional of the bare correlation functions $W_{B}$ to its
corresponding renormalised form $W_{R}$. \ To provide a convenient means of
handling the dependence of the correlation and response functions on the
dimensional parameters $\nu _{0}$ and $D_{0}$, we rescale ${\bf V}$ and $%
{\bf f}$ by introducing bare fields defined by 
\[
{\bf V}(\hat{k})=\left( \frac{D_{0}}{\nu _{0}^{3}}\right) ^{\frac{1}{2}}{\bf %
V}_{B}({\bf k},\omega _{B}), 
\]
and 
\[
{\bf f}(\hat{k})=\left( \frac{D_{0}}{\nu _{0}}\right) ^{\frac{1}{2}}{\bf f}%
_{B}({\bf k},\omega _{B}), 
\]
with bare frequency 
\[
\omega _{B}=\frac{\omega }{\nu _{0}}. 
\]
These bare fields preserve the form of the NS equations, apart from
explicitly introducing the non-dimensional coupling constant, defined by 
\begin{equation}
g_{0}=\frac{D_{0}}{6\pi ^{2}\nu _{0}^{3}},  \label{a37}
\end{equation}
in which the appropriateness of the numerical factor will appear later from
the loop-expansion of $W.$

Under the renormalisation (\ref{a28}), the bare fields are replaced by
renormalised fields, to which they are related by 
\[
{\bf V}_{B}({\bf k},\omega _{B})=\left( \frac{Z_{\nu }^{3}}{Z_{D}}\right) ^{%
\frac{1}{2}}{\bf V}_{R}({\bf k},\omega _{R}) 
\]
and 
\[
{\bf f}_{B}({\bf k},\omega _{B})=\left( \frac{Z_{\nu }}{Z_{D}}\right) ^{%
\frac{1}{2}}{\bf f}_{R}({\bf k},\omega _{R}), 
\]
where 
\[
\omega _{R}=\frac{\omega }{\nu }. 
\]
These relations follow from two requirements. First, the form of the NS
equations (\ref{a15}) must again be preserved, with the new constants $\nu $
and $D$ resulting in a renormalised coupling constant 
\begin{equation}
g=\frac{D}{6\pi ^{2}\nu ^{3}}.  \label{a41}
\end{equation}
Second, we have to satisfy the crucial requirement that ${\cal P}({\bf f}),$
as given in (\ref{a20}), remains invariant under renormalisation. Indeed,
satisfaction of these conditions implies the desired relation between $W_{B}$
and $W_{R}$, which from (\ref{a23}) and (\ref{a25}), is readily found to be 
\begin{equation}
W_{R}({\bf J},\widetilde{{\bf J}},t_{s})=W_{B}\left( \frac{1}{Z_{\nu }}%
\left( \frac{Z_{D}}{Z_{\nu }}\right) ^{1/2}{\bf J},\frac{1}{Z_{\upsilon }}%
\left( \frac{Z_{\upsilon }}{Z_{D}}\right) ^{1/2}\widetilde{{\bf J}},\,\frac{%
Z_{s}}{Z_{\nu }^{s}}\left( \frac{Z_{D}}{Z_{\nu }}\right) ^{s/2}t_{s}\right) .
\label{a42}
\end{equation}

The foregoing provides the basis of our calculation of $\zeta _{n}$, which
involves the following four stages. First, we use (\ref{a42}) and the
binomial expansion of (\ref{a1}) to develop a short distance expansion for $%
S_{n}(r),$ by substituting an operator product expansion (OPE) [7,8] for
each term, based on the operators (\ref{a9}). As shown in Section \ref
{section3}, this yields the scaling of $S_{n}(r)$ in terms of $uv$ fixed
point values of standard RG functions. The second stage of the calculation
is to demonstrate that the required $uv$ fixed point of the RG actually
exists, and then to deduce the corresponding fixed point coupling constant $%
g_{\ast }$. This is done in Section \ref{section4} by considering the
renormalisation of the linear response function, using the renormalised
functional in the form obtained from (\ref{a23}).\ The third stage is to
calculate the specific fixed point RG parameters which give the anomalous
component of $\zeta _{n}$. To do this, we have to consider the
renormalisation of appropriate nonlinear Green's functions involving the
composite operators defined in (\ref{a9}). These are identified and
evaluated in Section \ref{section5}. Having calculated the anomalous scaling
exponent $-\tau _{np}$ of the $p$th term in the binomial expansion of $%
S_{n}(r)$, in the fourth and final stage of the calculation, we derive a
simple algebraic expression for $\zeta _{n}$ by maximising \ $\tau _{np}$,
with respect to integer values of $p$, and subtracting this maximum from the
Kolmogorov value (\ref{a3}). The results obtained for $\zeta _{n}$ are
presented in Section \ref{section6}, where they are shown to be in good
agreement with experimental measurements at all orders for which reliable
data exists. Finally, the mathematical proofs, deferred during the
calculation of the exponents, are presented in Sections \ref{section7}-\ref
{section9}, and comprise: (a) the demonstration that sweeping effects can be
eliminated by means of a random Galilean transformation of the velocity
field; (b) the derivation of the modified quadratic form from the 1-loop
expansion; and (c) the derivation of the dominant terms of the OPEs.

\section{The Structure Function Expansions}

\label{section3}

In applying the OPE technique to (\ref{a1}) the first point to appreciate is
that the orders $n=2,$ $n=3$ and $n\geq 4$ require separate treatment. The
factor distinguishing $S_{2}$ and $S_{3}$ from the higher order $S_{n}$ is
that the latter involve composite operator products, whereas $S_{2}$ and $%
S_{3}$ do not. Also, $S_{3}$ is exceptional in representing a transition at
which corrections to the Kolmogorov exponents (\ref{a3}) change from
positive at $n=2$ to negative at $n\geqslant 4$, with no correction
occurring at $n=3$ in accordance with the known exact scaling law, which is
verified, within the present framework, in Section \ref{section8}. This sign
change is caused precisely because composite operator products appear in $%
S_{n}$ when $n\geqslant 4.$

We begin, therefore, with the relatively straightforward case of $S_{2}$.
According to (\ref{a1}), we have 
\begin{equation}
S_{2}(r)=2\left( \left\langle {\bf v}^{2}\right\rangle -\left\langle
v_{+}v_{-}\right\rangle \right) ,  \label{a43}
\end{equation}
which shows that the scaling of $S_{2}$ is determined by the behaviour of
the operator product $v_{+}v_{-}$ as $r\rightarrow 0$. The form of its OPE
is established in Section \ref{section9} after the necessary mathematical
apparatus has been set up. Its proof is given there to the accuracy of the
calculation, ie up to and including terms of order $g^{2}$. We shall show
that\ the operators which appear in its OPE are: (a) the unit operator $I$,
with constant coefficient $E/3$, where $E$ is the turbulence energy; (b) the
dominant longitudinal quadratic composite operator $O_{2}(\hat{x})$, which
gives the leading scaling behaviour; and (c) subdominant operators including
all transverse operators and the longitudinal higher order composite
operators $O_{s}(\hat{x})$. However, we shall only be concerned with the
dominant operators and so we write the expansion as \ 
\begin{equation}
v_{+}v_{-}=\frac{1}{3}EI+C_{2}(r)O_{2}(\hat{x})+\ldots ,  \label{a44}
\end{equation}
where the dots indicate the additional subdominant terms.The scaling
behaviour of this operator product can be found in the usual way from the RG
equation satisfied by the leading Wilson coefficient $C_{2}(r)$\ [12].

We start by considering an arbitrary equal time correlation function of
order $l$, given by 
\begin{equation}
H_{\alpha _{1}...\alpha _{l}}(\hat{x}_{1},...,\hat{x}_{l})=\left\langle
v_{\alpha _{1}}(\hat{x}_{1})...v_{\alpha _{l}}(\hat{x}_{l})\right\rangle .
\label{a45}
\end{equation}
\ If we insert (\ref{a44}) into this correlation function, we get 
\begin{equation}
H_{\alpha _{1}...\alpha _{l}}(\hat{x}_{1},...,\hat{x}_{l},\hat{x}+\frac{r}{2}%
\hat{\imath},\hat{x}-\frac{r}{2}\hat{\imath})=\frac{E}{3}H_{\alpha
_{1}...\alpha _{l}}(\hat{x}_{1},...,\hat{x}_{l})+C_{2}(r)Q_{\alpha
_{1}...\alpha _{l}}^{(2)}(\hat{x}_{1},...,\hat{x}_{l},\hat{x})+\ldots ,
\label{a46}
\end{equation}
where, in general, $Q_{\alpha _{1}...\alpha _{l}}^{(s)}$\ is the inserted
correlation function defined by

\begin{equation}
Q_{\alpha _{1}...\alpha _{l}}^{(s)}\left( \hat{x}_{1},...,\hat{x}_{l},\hat{x}%
\right) =\left\langle v_{\alpha _{1}}(\hat{x}_{1})...v_{\alpha _{l}}(\hat{x}%
_{l})O_{s}(\hat{x})\right\rangle ,  \label{a47}
\end{equation}
and $\hat{\imath}$ is a unit vector along the $x$-axis.

We can deduce the RG equation satisfied by the Wilson coefficient $C_{2}$ in
(\ref{a44}) from (\ref{a46}), given the RG equations satisfied by $H_{\alpha
_{1}...\alpha _{l}}$ and $Q_{\alpha _{1}...\alpha _{l}}^{(2)}$. To obtain
the latter, we need the transformation rule for the equal time generator of
these correlation functions, which we denote by $W^{(e)}({\bf J,}t_{s}$%
).This follows in a straightforward manner by taking time independent
sources in (\ref{a42}), and integrating with respect to $\omega _{B}$ and $%
\omega _{R}$, with the $\widetilde{{\bf J}\text{ }}$dependence, which is
irrelevant here, suppressed. To simplify the result, we shall anticipate the
fact, which we demonstrate in Section \ref{section4}, that 
\begin{equation}
Z_{D}=Z_{\nu }.  \label{a48}
\end{equation}
We then get 
\begin{equation}
W_{R}^{(e)}({\bf J},t_{s})=W_{B}^{(e)}\left( {\bf J},Z_{s}t_{s}\right) .
\label{a49}
\end{equation}

According to this relation, the bare and renormalised forms of $H_{\alpha
_{1}...\alpha _{l}}$ are equal. Hence, when we change the renormalisation
scale, which we denote by $\mu $, the Fourier transform of $H_{\alpha
_{1}...\alpha _{l}}$\ changes according to the RG equation 
\begin{equation}
{\cal D}H_{\alpha _{1}...\alpha _{l}}=0,  \label{a50}
\end{equation}
where ${\cal D}$ is the standard RG operator defined by 
\begin{equation}
{\cal D}=\mu \frac{\partial }{\partial \mu }+\beta (g)\frac{\partial }{%
\partial g},  \label{a51}
\end{equation}
with 
\begin{equation}
\beta (g)=\mu \frac{\text{d}g}{\text{d}\mu }.  \label{a52}
\end{equation}
In the case of $Q_{\alpha _{1}\ldots \alpha _{l}}^{(s)}$, we obtain from (%
\ref{a47}) and (\ref{a49}) the relation 
\[
(Q_{\alpha _{1}...\alpha _{l}}^{(s)})_{R}=Z_{s}(Q_{\alpha _{1}...\alpha
_{l}}^{(s)})_{B}, 
\]
which leads to the RG equation 
\begin{equation}
{\cal D}Q_{\alpha _{1}...\alpha _{l}}^{(s)}=\gamma _{s}Q_{\alpha
_{1}...\alpha _{l}}^{(s)},  \label{a54}
\end{equation}
where $\gamma _{s}$ is the anomalous dimension of $O_{s}$ given by 
\begin{equation}
\gamma _{s}=\mu \frac{\text{d}}{\text{d}\mu }\log Z_{s}.  \label{a55}
\end{equation}
For ease of notation, we have dropped the suffix $R$ in the RG equations (%
\ref{a50}) and (\ref{a54}), since we shall always be dealing with relations
between renormalised functions.

We now apply the RG operator (\ref{a51}) to the Fourier transform of (\ref
{a46}), and make use of (\ref{a50}) and (\ref{a54}), to get 
\begin{equation}
0=\left( {\cal D}C_{2}+\gamma _{2}C_{2}\right) Q_{\alpha _{1}...\alpha
_{l}}^{(2)}+\ldots \text{ \ .}  \label{a186}
\end{equation}
As this equation holds for arbitrary $Q_{\alpha _{1}...\alpha _{l}}^{(2)}$,
it follows that 
\begin{equation}
{\cal D}C_{2}=-\gamma _{2}C_{2}.  \label{a188}
\end{equation}
which is the RG equation satisfied by the Wilson coefficients in (\ref{a44}).

The standard solution of this equation, corresponding to an {\it uv }fixed
point [12], now gives for the leading term of (\ref{a44}) the scaling
behaviour 
\begin{equation}
C_{2}(r)\sim r^{2/3-\gamma _{2}^{\ast }},  \label{a189}
\end{equation}
where the star denotes the fixed point value of (\ref{a55}). This result, in
conjunction with (\ref{a43}) and (\ref{a44}), yields the scaling exponent
for $S_{2}(r),$namely 
\begin{equation}
\zeta _{2}=\frac{2}{3}+\Delta _{2},  \label{a59}
\end{equation}
where 
\begin{equation}
\Delta _{2}=-\gamma _{2}^{\ast }.  \label{a60}
\end{equation}
We shall calculate $\Delta _{2}$ in Section \ref{section5}.

Consider now the general case for even orders $n=2m>2.$ Introducing the
general composite operator product 
\begin{equation}
\Lambda _{ss^{\prime }}(\hat{x},r)=O_{s}\left( \hat{x}+\frac{r}{2}\hat{\imath%
}\right) O_{s^{\prime }}\left( \hat{x}-\frac{r}{2}\hat{\imath}\right) ,
\label{a68}
\end{equation}
and taking advantage of the isotropic symmetry, we can write the binomial
expansion of (\ref{a1}) as 
\begin{equation}
S_{n}\left( r\right) =n!\left\langle 2\sum_{p=0}^{m-1}\left( -\right)
^{p}\Lambda _{n-p,p}\left( \hat{x},r\right) +\left( -\right) ^{m}\Lambda
_{m,m}\left( \hat{x},r\right) \right\rangle .  \label{a69}
\end{equation}

We can identify the dominant term of the OPE of $\Lambda _{n-p,p}$ by
factoring out the product $(v_{+}v_{-})^{p}$ and using the fact that, by (%
\ref{a44}), its expansion begins with the unit operator. We will justify
this process in Section \ref{section9}. This implies that the OPE of $%
\Lambda _{n-p,p}$ itself takes the form 
\begin{equation}
\Lambda _{n-p,p}(\hat{x},r)=C_{p,m-p}(r)O_{2(m-p)}(\hat{x})+\ldots ,
\label{a70}
\end{equation}
where again the dots indicate subdominant terms.Substituting (\ref{a70}) in (%
\ref{a69}), we get 
\begin{equation}
S_{n}(r)=\,n!\left\{ 2\sum_{p=0}^{m-1}(-)^{p}C_{p,m-p}(r)\left\langle
O_{2(m-p)}(\hat{x})\right\rangle +(-)^{m}C_{m,m}(r)\right\} +\ldots ,
\label{a72}
\end{equation}
the averages of the composite operators being independent of $\hat{x}$ for
homogeneous isotropic turbulence.

To find $\zeta _{n}$ from this expansion, we have to determine which term or
terms on the right hand side yield the negative correction of maximum
magnitude to $\zeta _{n}^{Kol}$. As before, this is deduced from the RG
equation for the Wilson coefficient $C_{p,s}$, which we derive next.

We begin by inserting (\ref{a68}) into the general correlation function (\ref
{a45}) to obtain the general inserted correlation function 
\begin{equation}
R_{\alpha _{1}...\alpha _{l}}^{(ss^{\prime })}(\hat{x}_{1},...,\hat{x}_{l},%
\hat{x}+\frac{1}{2}r\hat{\imath},\hat{x}-\frac{1}{2}r\hat{\imath}%
)=\left\langle v_{\alpha _{1}}(\hat{x}_{1})...v_{\alpha _{l}}(\hat{x}%
_{l})\Lambda _{ss^{\prime }}(\hat{x},r)\right\rangle .  \label{a73}
\end{equation}
According to (\ref{a49}) its bare and renormalised forms are connected by
the relation 
\[
\left( R_{\alpha _{1}...\alpha _{l}}^{(ss^{\prime })}\right)
_{R}=Z_{s}Z_{s^{\prime }}\left( R_{\alpha _{1}...\alpha _{l}}^{(ss^{\prime
})}\right) _{B}, 
\]
from which it follows that $R_{\alpha _{1}...\alpha _{l}}^{(ss^{\prime })}$
satisfies the RG equation 
\begin{equation}
{\cal D}R_{\alpha _{1}...\alpha _{l}}^{(ss^{\prime })}=\left( \gamma
_{s}+\gamma _{s^{\prime }}\right) R_{\alpha _{1}...\alpha _{l}}^{(ss^{\prime
})}.  \label{a75}
\end{equation}

Next, we insert the expansion (\ref{a70}) into the general correlation
function (\ref{a45}), and use the definitions (\ref{a47}) and (\ref{a73}),
to get 
\[
R_{\alpha _{1}...\alpha _{l}}^{(n-p,p)}\left( \hat{x}_{1},...,\hat{x}_{l},%
\hat{x}+\frac{r}{2}\hat{\imath},\hat{x}-\frac{r}{2}\hat{\imath}\right)
=C_{p,m-p}\left( r\right) Q_{\alpha _{1}...\alpha _{l}}^{(2s)}\left( \hat{x}%
_{1},...,\hat{x}_{l},\hat{x}\right) . 
\]
We then apply the RG operator (\ref{a51}) to the Fourier transform of this
equation, and substitute (\ref{a54}) and (\ref{a75}) to obtain 
\[
\ Q_{\alpha _{1}...\alpha _{l}}^{2(m-p)}\left\{ {\cal D}C_{p,m-p}+\left(
\gamma _{2(m-p)}-\gamma _{p}-\gamma _{n-p}\right) C_{p,m-p}\right\} +\ldots
=0, 
\]
from which it follows that 
\begin{equation}
{\cal D}C_{p,m-p}=-\left( \gamma _{2(m-p)}-\gamma _{p}-\gamma _{n-p}\right)
C_{p,m-p}.  \label{a78}
\end{equation}

We now invoke the standard solution of (\ref{a78}), applicable at the {\it %
uv }fixed point [12], to obtain the scaling relation 
\begin{equation}
C_{p,m-p}(r)\sim r^{n/3-\tau _{np}},  \label{a79}
\end{equation}
where 
\begin{equation}
\tau _{np}=\gamma _{2(m-p)}^{\ast }-\gamma _{p}^{\ast }-\gamma _{n-p}^{\ast
}.  \label{a80}
\end{equation}
Upon substituting (\ref{a79}) in (\ref{a72}), it is immediately evident that
the scaling exponent of $S_{n}(r)$ is given by 
\begin{equation}
\zeta _{n}=\frac{n}{3}-\tau _{n},  \label{a81}
\end{equation}
where 
\begin{equation}
\tau _{n}=\max_{p}\text{ }\tau _{np},\text{ \ \ \ \ \ for \ \ \ \ \ \ }%
n=2m>2.  \label{a82}
\end{equation}
Once $\gamma _{s}^{\ast }$ has been evaluated from (\ref{a55}), at the fixed
point, which we do in Section \ref{section5}, it is a simple matter to
evaluate $\tau _{n}$, as we show in Section \ref{section6}.

Odd orders with $n=2m+1>3$ may be treated similarly with minor adjustments
to allow for the fact that the expansions involve odd powers. In this case,
however, it is immediately evident that the dominant scaling must arise from
the Wilson coefficient of the unit operator corresponding to $p=m,$ because
averaging wipes out other terms by virtue of the fact that $\left\langle
O_{2s+1}\right\rangle =0.$ Hence, we obtain 
\begin{equation}
\tau _{n}=-\left( \gamma _{m}^{\ast }+\gamma _{m+1}^{\ast }\right) \text{ \
\ \ \ \ \ for \ \ \ \ \ \ \ }n=2m+1>3.  \label{a83}
\end{equation}
Again, the justification of the relevant expansions is given in Section \ref
{section9}.

\section{The Linear Response}

\label{section4}

In order to evaluate $\tau _{n}$, we have to establish that an {\it uv}
fixed point exists, which entails showing that the RG $\beta $ function (\ref
{a52}) possesses a zero 
\begin{equation}
\beta (g_{\ast })=0,  \label{a84}
\end{equation}
at which 
\begin{equation}
\text{d}\beta /\text{d}g<0.  \label{a85}
\end{equation}
To do this we must first determine the dependence of the renormalisation
constants $Z_{\nu }$ and $Z_{D\text{ }}$on the renormalisation scale $\mu $.
We will then verify that (\ref{a48}) holds and use this fact to calculate $%
g_{\ast }$ from $Z_{\nu }.$

Consider $Z_{\nu }$. According to the general theory of renormalisation
[14], we have an expansion of the form 
\begin{equation}
Z_{\nu }=1+g\,a_{1\nu }\log \left( \frac{\mu }{\kappa }\right) +g^{2}\left\{ 
\frac{a_{1\nu }^{2}}{2}\log ^{2}\left( \frac{\mu }{\kappa }\right) +a_{2\nu
}\log \left( \frac{\mu }{\kappa }\right) \right\} +...\text{ \ .}
\label{a86}
\end{equation}
Here $\kappa $ is the wavenumber cut-off which provides the intermediate
regulation of the divergent integrals.This is an {\it ir} wavenumber of the
order of $L^{-1}$, where $L$ is the typical length scale of the large scale
flow. Divergences arise in the limit $\kappa \rightarrow 0,$ corresponding
to the inertial range limit $r/L\rightarrow 0.$ The constants $a_{1\nu }$
and $a_{2\nu }$ will be calculated by eliminating the logarithmic
divergences, at 1 and 2-loop orders respectively, from the 1PI Green's
function $\Gamma _{\alpha \beta }(\hat{k},\hat{l}),$which is the inverse of
the Fourier transform $G_{\alpha \beta }(\hat{k},\hat{l})$ of the linear
response function\ 
\begin{equation}
G_{\alpha \beta }\left( \hat{x},\hat{x}^{\prime }\right) =\left\langle \frac{%
\delta v_{\alpha }\left( \hat{x}\right) }{\delta f_{\beta }\left( \hat{x}%
^{\prime }\right) }\right\rangle .  \label{a87}
\end{equation}

$\Gamma _{\alpha \beta }$, and the other 1PI functions that we shall
require, are generated from the functional $K$, which is obtained in the
usual way by performing a Legendre transformation on $W_{c}=\log W$, with
respect to the sources of the elementary fields, ${\bf J}$ and $\widetilde{%
{\bf J}}$, while holding the composite operator sources $t_{s}$ fixed
[7,13]. The new source fields for $K$ are therefore given by

\[
{\bf u}(\hat{k})=\left( 2\pi \right) ^{4}\frac{\delta W_{c}}{i\delta {\bf J}%
(-\hat{k})}, 
\]
and\ \ 
\[
\widetilde{{\bf u}}(\hat{k})=(2\pi )^{4}\frac{\delta W_{c}}{i\delta 
\widetilde{{\bf J}}(-\hat{k})}, 
\]
with $K$ itself given in terms of its source fields by 
\[
K({\bf u},\widetilde{{\bf u}},t_{s})=-W_{c}+i\int \left\{ {\bf J}(-\hat{k}%
)\cdot {\bf u}(\hat{k})+\widetilde{{\bf J}}(-\hat{k})\cdot \widetilde{{\bf u}%
}(\hat{k})\right\} D\hat{k}. 
\]

It follows, therefore, that 
\[
\Gamma _{\alpha \beta }(\hat{k},\hat{l})=(2\pi )^{8}\frac{\delta ^{2}K}{%
i\delta \widetilde{u}_{\alpha }(\hat{k})\delta u_{\beta }(\hat{l})}. 
\]
Introduction of the reduced forms 
\begin{equation}
G_{\alpha \beta }(\hat{k},\hat{l})=(2\pi )^{4}\delta (\hat{k}+\hat{l}%
)P_{\alpha \beta }({\bf k})G(\hat{k}),  \label{a92}
\end{equation}
and 
\[
\Gamma _{\alpha \beta }(\hat{k},\hat{l})=(2\pi )^{4}\delta (\hat{k}+\hat{l}%
)P_{\alpha \beta }({\bf k})\Gamma (\hat{k}), 
\]
then leads to the standard relation 
\begin{equation}
\Gamma (\hat{k})=G(\hat{k})^{-1}.  \label{a94}
\end{equation}
We can now use ( \ref{a42}) to show that the connection between the bare and
renormalised forms is 
\begin{equation}
\Gamma _{R}=Z_{\nu }\Gamma _{B},  \label{a95}
\end{equation}
which demonstrates the suitability of $\Gamma (\hat{k})$ as a basis for
determining $Z_{\nu }$.

In carrying out the renormalisation of $\Gamma (\hat{k})$ to obtain the
coefficients in (\ref{a86}), we choose the normalisation point to be $\hat{k}%
=\hat{m}$, where 
\[
\hat{m}=({\bf m},\omega _{m}=0). 
\]
Here ${\bf m}$ is a fixed vector of magnitude 
\[
\left| {\bf m}\right| =\mu , 
\]
the direction of which need not be specified, because the geometrical factor
is contained in $P_{\alpha \beta }({\bf m)}$ which cancels off. The
expansion (\ref{a86}) is used in conjunction with a normalisation condition
that sets $\Gamma (\hat{m})$ equal to its tree level value. Thus, from (\ref
{a16}) and (\ref{a94}), we have 
\begin{equation}
\Gamma (\hat{m})=\Gamma _{0}(\hat{m})=G_{0}(\hat{m})^{-1}=\tau (\mu )^{-1},
\label{a210}
\end{equation}
and so the 1-loop term satisfies the normalisation condition 
\begin{equation}
\Gamma _{1}(\hat{m})=0.  \label{a99}
\end{equation}

The Feynman diagram giving the 1-loop term of $\Gamma _{\alpha \beta }(\hat{m%
})$ is shown in Fig.3({\it i}). The standard rules apply to such diagrams
with the following assignments, which are shown in Fig.1:

\begin{enumerate}
\item  External lines represent functional differentiation with respect to $%
{\bf u}(\hat{k})$ when continuous, and $\widetilde{{\bf u}}{\bf (}\hat{k})$,
when dotted.The diagram is divided by a factor of $i$ for each
differentiation with respect to $\widetilde{{\bf u}\text{.}}$

\item  A continuous line linking two vertices denotes the reduced velocity
correlation function defined through 
\[
\left\langle v_{\alpha }(\hat{k})v_{\beta }(\hat{l})\right\rangle =(2\pi
)^{4}\delta (\hat{k}+\hat{l})Q_{\alpha \beta }(\hat{k}), 
\]
\ and given by 
\[
Q_{\alpha \beta }(\hat{k})=D(k)\left| G(\hat{k})\right| ^{2}P_{\alpha \beta
}\left( {\bf k}\right) . 
\]
For ease of notation, we omit zero-order labels in writing down mathematical
expressions for the diagrams.

\item  A half dotted/half continuous line connecting two vertices represents 
$i$ times the zero-order response function 
\[
G_{\alpha \beta }(\hat{k})=G(\hat{k})P_{\alpha \beta }\left( {\bf k}\right)
. 
\]

\item  The NS vertex with one dotted and two continuous lines represents $%
P_{\alpha \beta \gamma }({\bf k}),$ the argument of which is associated with
the dotted leg, with ${\bf k}$ directed away from the node.
\end{enumerate}

Returning now to the 1-loop diagram for $\Gamma _{\alpha \beta }(\hat{m})$,
we note that it has a symmetry factor of 1. Hence, it yields a contribution
to $\Gamma _{1}(\hat{m})$ given by 
\[
P_{\alpha \beta }({\bf m})\Gamma _{1}^{\prime }(\hat{m})=\int D\hat{p}%
\,P_{\alpha \gamma \delta }({\bf m})P_{\lambda \nu \beta }({\bf m}-{\bf p}%
)G_{\gamma \lambda }(\hat{m}-\hat{p})Q_{\delta \nu }(\hat{p}). 
\]
We can extract the logarithmic divergence from this integral by expanding
its integrand in powers of $p/m$. This is possible because the divergence
emanates from the region $p\sim \kappa $, while $\kappa \ll \mu $. A simple
calculation leads to 
\begin{equation}
\Gamma _{1}^{\prime }(\hat{m})=\frac{3}{2}g\tau (\mu )^{-1}I_{0}(\varepsilon
),  \label{a203}
\end{equation}
where 
\begin{equation}
I_{0}(\varepsilon )=\int\limits_{\varepsilon }^{\infty }\frac{dx}{x^{2}(x+1)}%
,  \label{a300}
\end{equation}
in which the lower limit of integration is 
\[
\varepsilon =\frac{\tau (\mu )}{\tau (\kappa )}. 
\]
Extracting the logarithmic singularity from this integral gives 
\begin{equation}
\Gamma _{1}^{\prime }(\hat{m})=-\tau (\mu )^{-1}g\log \left( \frac{\mu }{%
\kappa }\right) .  \label{a104}
\end{equation}
To this we have to add the term arising from the counterterm vertex shown in
Fig.2({\it i}). This contributes the term $P_{\alpha \beta }\left( {\bf m}%
\right) \Gamma _{1}^{\prime \prime }(\hat{m})$ where 
\begin{equation}
\Gamma _{1}^{\prime \prime }(\hat{m})=\Delta Z_{\nu }\tau (\mu
)^{-1}=a_{1\nu }\tau \left( \mu \right) ^{-1}g\log \left( \frac{\mu }{\kappa 
}\right) .  \label{a105}
\end{equation}

But, from the normalisation condition (\ref{a99}), we have 
\[
\Gamma _{1}^{\prime }(\hat{m})+\Gamma _{1}^{\prime \prime }(\hat{m})=0, 
\]
which, upon substituting (\ref{a104}) and (\ref{a105}), yields 
\[
a_{1\nu }=1. 
\]

We next carry out the analogous calculation for $Z_{D}$ and show that its
corresponding coefficient $a_{1D}$ also equals 1, thereby verifying that the
condition (\ref{a48}) is satisfied at 1-loop order. Here the relevant 1PI
function is the correlation function given by 
\[
\Pi _{\alpha \beta }(\hat{k},\hat{l})=\left( 2\pi \right) ^{8}\frac{\delta
^{2}K}{i\delta \widetilde{u}_{\alpha }i\delta \widetilde{u}_{\beta }}, 
\]
which is readily shown to be related to the velocity correlation function $%
Q_{\alpha \beta }(\hat{k},\hat{l})$ by [13] 
\[
\Pi _{\alpha \beta }(\hat{k},\hat{l})=\int \Gamma _{\alpha \lambda }(\hat{k},%
\hat{p})\Gamma _{\beta \mu }(\hat{l},\hat{q})Q_{\lambda \mu }(\hat{p},\hat{q}%
)\,d\hat{p}d\hat{q}\,. 
\]

Substituting the reduced forms 
\[
\Pi _{\alpha \beta }(\hat{k},\hat{l})=(2\pi )^{4}\delta (\hat{k}+\hat{l}%
)P_{\alpha \beta }({\bf k})\Pi (\hat{k}), 
\]
and 
\[
Q_{\alpha \beta }(\hat{k},\hat{l})=(2\pi )^{4}\delta (\hat{k}+\hat{l}%
)\,P_{\alpha \beta }({\bf k})Q(\hat{k}), 
\]
we get 
\[
\Pi (\hat{k})=\left| \Gamma (\hat{k})\right| ^{2}Q(\hat{k}). 
\]
From this result and (\ref{a42}) and (\ref{a94}), we find that the bare and
renormalised forms of $\Pi $ are related by 
\[
\Pi _{R}=Z_{D}\Pi _{B}, 
\]
which confirms that $\Pi (\hat{k})$ is the appropriate 1PI function to use
for calculating $Z_{D}$.

The normalisation condition is again chosen to be consistent with the tree
level approximation. That is, we set\ 
\[
\Pi (\hat{m})=\Pi _{0}(\hat{m})=D\left( \mu \right) , 
\]
so that the 1-loop term satisfies the normalisation condition 
\begin{equation}
\Pi _{1}(\hat{m})=0.  \label{a115}
\end{equation}

The 1-loop Feynman diagram for $\Pi _{\alpha \beta }(\hat{m})$ is shown in
Fig.3({\it ii}). It has a symmetry factor of 1/2, and makes a contribution
to $\Pi _{1}(\hat{m})$ which is given by 
\[
P_{\alpha \beta }({\bf m})\Pi _{1}^{\prime }(\hat{m})=\frac{1}{2}\int D\hat{p%
}\,P_{\alpha \gamma \delta }({\bf m})P_{\beta \lambda \nu }({\bf m}%
)Q_{\gamma \lambda }(\hat{p})Q_{\delta \nu }(\hat{m}-\hat{p}). 
\]
In extracting the logarithmic singularity from this integral, we must take
into account the fact that the symmetry of the integrand results in
singularities of equal strength at both $p\sim \kappa $, and $\left| {\bf p-m%
}\right| \sim \kappa $, the effect of which compensates for the symmetry
factor. Consequently, we get 
\begin{equation}
\Pi _{1}^{\prime }(\hat{m})=-D(\mu )g\log \left( \frac{\mu }{\kappa }\right)
.  \label{a117}
\end{equation}
The $Z_{D}$ counterterm, which is shown in Fig.2({\it ii}), contributes a
term to $\Pi _{1}(\hat{m})$ given by 
\begin{equation}
\Pi _{1}^{\prime \prime }(\hat{m})=\Delta Z_{D}D\left( \mu \right)
=a_{1D}D\left( \mu \right) g\log \left( \frac{\mu }{\kappa }\right) .
\label{a118}
\end{equation}
But, from the normalisation condition (\ref{a115}), we have 
\[
\Pi _{1}^{\prime }(\hat{m})+\Pi _{1}^{\prime \prime }(\hat{m})=0, 
\]
and substitution of (\ref{a117}) and (\ref{a118}) leads to 
\[
a_{1D}=1. 
\]

We shall take the equality of the 1-loop coefficients of $Z_{\nu }$ and $%
Z_{D}$ as establishing that (\ref{a48}) holds. This allows us to calculate
the {\it uv} fixed point from the linear response function alone as
follows.We use the standard result [14] 
\begin{equation}
\beta (g)=-g\,\mu \frac{\partial }{\partial \mu }\log Z_{g}\left( 1+g\,\frac{%
\partial }{\partial g}\log Z_{g}\right) ^{-1},  \label{a121}
\end{equation}
where $Z_{g}$ is the renormalisation constant associated with the coupling
constant.From the definition $Z_{g}=g_{0}/g$ and (\ref{a28}),(\ref{a37}),(%
\ref{a41}) and (\ref{a48}), we get 
\[
Z_{g}=Z_{\nu }^{-2}. 
\]
Inserting this result in (\ref{a121}) and substituting the expansion (\ref
{a86}), leads to 
\[
\beta \left( g\right) =2g^{2}(1+a_{2\nu }g). 
\]
This yields an {\it uv} fixed point 
\begin{equation}
g_{\ast }=-\frac{1}{a_{2\nu }},  \label{a124}
\end{equation}
which satisfies (\ref{a84}) and (\ref{a85}) provided that $a_{2\nu }<0$. It
remains,then, to calculate $a_{2\nu }.$

The constant $a_{2\nu }$ is obtained from the 2-loop term of $\Gamma
_{\alpha \beta }(\hat{k})$, namely $P_{\alpha \beta }({\bf m})\Gamma _{2}(%
\hat{k})$. At the normalisation point it must satisfy the condition 
\begin{equation}
\Gamma _{2}(\hat{m})=0,  \label{a125}
\end{equation}
by virtue of (\ref{a68}). Only two Feynman diagrams yield logarithmic
divergences. They are shown in Figs.3({\it iii}) and ({\it iv}).They
contribute the terms 
\begin{eqnarray*}
P_{\alpha \beta }({\bf m})\Gamma _{2}^{\prime }(\hat{m}) &=&-\int D\hat{p}\,D%
\hat{q}Q_{\delta \epsilon }(\hat{p})Q_{\lambda \rho }(\hat{q}) \\
&&\times G_{\kappa \gamma }(\hat{m}-\hat{p})G_{\nu \sigma }(\hat{m}-\hat{p}-%
\hat{q})G_{\tau \mu }(\hat{m}-\hat{p}) \\
&&\times P_{\alpha \gamma \delta }({\bf m})P_{\mu \beta \epsilon }({\bf m}-%
{\bf p})P_{\kappa \lambda \nu }({\bf m}-{\bf p})P_{\varrho \sigma \tau }(%
{\bf m}-{\bf p}-{\bf q}),
\end{eqnarray*}
and 
\begin{eqnarray}
P_{\alpha \beta }({\bf m})\Gamma _{2}^{\prime \prime }(\hat{m}) &=&-\int D%
\hat{p}\,D\hat{q}Q_{\delta \sigma }(\hat{p})Q_{\lambda \mu }(\hat{q})
\label{a127} \\
&&\times G_{\kappa \gamma }(\hat{m}-\hat{p})G_{\varrho \nu }(\hat{m}-\hat{p}-%
\hat{q})G_{\epsilon \tau }(\hat{m}-\hat{q})  \nonumber \\
&&\times P_{\alpha \gamma \delta }({\bf m})P_{\epsilon \mu \beta }({\bf m}-%
{\bf q})P_{\kappa \lambda \nu }({\bf m}-{\bf p})P_{\varrho \sigma \tau }(%
{\bf m}-{\bf p}-{\bf q}).  \nonumber
\end{eqnarray}

To extract the logarithmic singularities from these integrals, we expand
their integrands in powers of both $p/m$ and $q/m$. We do this in two steps.
First, we integrate over frequencies and solid angles to get 
\begin{equation}
\Gamma _{2}^{\prime }(\hat{m})=-\frac{9}{4}g^{2}\tau (\mu )^{-1}I_{1}\left(
\varepsilon \right) ,  \label{a208}
\end{equation}
and 
\begin{equation}
\Gamma _{2}^{\prime \prime }(\hat{m})=-\frac{9}{4}g^{2}\tau (\mu
)^{-1}I_{2}\left( \varepsilon \right) ,  \label{a205}
\end{equation}
where 
\begin{equation}
I_{1}\left( \varepsilon \right) =\int\limits_{\varepsilon }^{\infty
}\int\limits_{\varepsilon }^{\infty }\frac{dxdy}{x^{2}y^{2}\left( 1+x\right)
^{2}\left( 1+x+y\right) },  \label{a301}
\end{equation}
and 
\begin{equation}
I_{2}\left( \varepsilon \right) =\int\limits_{\varepsilon }^{\infty
}\int\limits_{\varepsilon }^{\infty }\frac{dxdy}{x^{2}y^{2}\left( 1+x\right)
\left( 1+y\right) \left( 1+x+y\right) }.  \label{a302}
\end{equation}
Secondly, we expand these double integrals for small $\varepsilon $ to
obtain 
\begin{equation}
\Gamma _{2}^{\prime }(\hat{m})=8g^{2}\tau (\mu )^{-1}\log \left( \frac{\mu }{%
\kappa }\right) ,  \label{a133}
\end{equation}
and 
\begin{equation}
\Gamma _{2}^{\prime \prime }(\hat{m})=\frac{21}{2}\tau (\mu )^{-1}g^{2}\log
\left( \frac{\mu }{\kappa }\right) .  \label{a134}
\end{equation}

To these two contributions to $\Gamma _{2}(\hat{m})$, we must add the
counterterm, which, by analogy with (\ref{a105}), takes the form 
\begin{equation}
\Gamma _{2}^{\prime \prime \prime }(\hat{m})=a_{2\nu }g^{2}\tau (\mu
)^{-1}\log \left( \frac{\mu }{\kappa }\right) .  \label{a135}
\end{equation}
Thus, the normalisation condition (\ref{a125}) becomes 
\[
\Gamma _{2}^{\prime }(\hat{m})+\Gamma _{2}^{\prime \prime }(\hat{m})+\Gamma
_{2}^{\prime \prime \prime }(\hat{m})=0, 
\]
which, by (\ref{a133})-(\ref{a135}), yields 
\[
a_{2\nu }=-\frac{37}{2}. 
\]
Therefore, from (\ref{a124}), we obtain the fixed point coupling constant 
\begin{equation}
g_{\ast }=\frac{2}{37},  \label{a138}
\end{equation}
which verifies that the residual coupling can be treated as weak.

Finally, we explain why the 2-loop topologies, which have been discarded in
calculating $\Gamma $, do not contribute to $a_{2\nu }.$As we shall explain
further in Section\ref{section7}, divergences arise in these diagrams when
it is possible for one or more soft wavevectors (ie values of $p$ and/or $%
q\ll m$ ) to flow through a correlator.However, if this entails the flow of
some or all of these wavevectors through the active (ie dotted) leg of the
NS vertex, then the logarithmic divergence will be suppressed by the extra
powers of $p$ and/or $q$. In the case of the 2-loop diagrams which we have
just calculated, the external hard wavevector $\hat{m}$ flows through the
active legs of all vertices, so no suppression occurs. However, in the case
of the remaining topologies at least one soft wavevector flowing through a
correlator must also flow through the active leg of a NS vertex. In the case
of the four remaining two loop topologies containing vertex corrections, the
logarithmic divergence is suppressed individually for each diagram, after
integration over the solid angles. In the case of the three remaining 2-loop
diagrams containing insertions of the 1-loop diagrams ({\it i) }and ({\it iv}%
) of Fig.3, suppression results after integrating over the solid angles and
summing over the diagrams, the overall cancellation being related to the
fact that the coefficients $a_{1\nu }$ and $a_{1D}$ associated with the two
types of insertion are equal. Likewise the four 1-loop diagrams containing
the counterterm vertices yield no net contribution to $a_{2\nu }$. The
treatment of the power and power$\times $logarithmic divergences arising in
integrals like (\ref{a300}),(\ref{a301}) and (\ref{a302}) is given in
Section \ref{section7}. For the moment we discard them because they are not
directly relevant to the actual calculation of the scaling exponents for
reasons already given.

\section{The Nonlinear Response}

\label{section5}

Having established that an {\it uv} fixed point exists, we can proceed with
the calculation of the anomalous dimension $\gamma _{s\text{ }}$of the
general operator $O_{s}(\hat{x})$, which is required for the evaluation of
the anomaly $\tau _{n}.$To do this in the simplest possible way, we must
identify a 1PI response function which can be renormalised by means of $%
Z_{s}.$ Elimination of the logarithmic divergences from such a function will
then enable us to determine the constants in the expansion 
\begin{equation}
Z_{s}=1+g_{\ast }a_{1}^{(s)}\log \left( \frac{\mu }{\kappa }\right) +g_{\ast
}^{2}\left\{ \frac{1}{2}\left( a_{1}^{(s)}\right) ^{2}\log ^{2}\left( \frac{%
\mu }{\kappa }\right) +a_{2}^{(s)}\log \left( \frac{\mu }{\kappa }\right)
\right\} +\ldots .  \label{a139}
\end{equation}
so that we can calculate $\gamma _{s}$ using (\ref{a55}).

Consider first the case $s=2$. Obviously, the required function must involve 
$O_{2}(\hat{x})$, which is the composite operator associated with the
longitudinal turbulence energy. In addition, it must involve the dynamic
response operator (\ref{a11}) in order to relate the anomaly $\tau _{2}$ to
the dynamics of the turbulence. This suggests that we should consider how
the turbulence energy responds on average to a change in the forcing.
Clearly, we can characterise the response of the turbulence energy at a
point $\hat{x}$ to a change in the forcing at two points $\hat{x}^{\prime }$
and $\hat{x}^{\prime \prime }$ by means of the nonlinear Green's function 
\begin{equation}
G_{\alpha \beta }^{(2)}(\hat{x}^{\prime },\hat{x}^{\prime \prime },\hat{x}%
)=\left\langle \frac{\delta ^{2}}{\delta f_{\alpha }(\hat{x}^{\prime
})\delta f_{\beta }(\hat{x}^{\prime \prime })}\left( \frac{v_{1}(\hat{x})^{2}%
}{2}\right) \right\rangle .  \label{a140}
\end{equation}
But the complexity of this object is such that its logarithmic divergences
cannot be summed using the renormalisation group in terms of the $Z_{2}$ and 
$Z_{\nu }$ counterterms alone. On the other hand, its average $\overline{G}%
_{\alpha \beta }^{(2)}(\hat{x})$ taken over the forcing separation $\hat{x}%
^{\prime }-\hat{x}^{\prime \prime }$, which gives a mean response to forcing
at the centroid of the excitation points, can be, as we shall show shortly.
Hence, its corresponding 1PI function provides a direct means of obtaining
the expansion (\ref{a139}) and so it provides an adequate basis for the \
calculation of $\gamma _{2}$.

This 1PI function is obtained as follows. We start with the Fourier
transform of (\ref{a140}), the reduced form of which is given by 
\begin{equation}
G_{\alpha \beta }^{(2)}(\hat{k},\hat{l},\hat{p})=(2\pi )^{4}\delta (\hat{k}+%
\hat{l}+\hat{p})G_{\alpha \beta }^{(2)}(\hat{k},\hat{l}),  \label{a141}
\end{equation}
where 
\begin{equation}
G_{\alpha \beta }^{(2)}(\hat{k},\hat{l})=P_{\alpha 1}\left( {\bf k}\right)
P_{\beta 1}\left( {\bf l}\right) G_{2}(\hat{k},\hat{l}).  \label{a142}
\end{equation}
Its corresponding 1PI response function follows from 
\begin{equation}
\Theta _{\alpha \beta }^{(2)}(\hat{k},\hat{l},\hat{p})=(2\pi )^{12}\frac{%
\delta ^{3}K}{\delta u_{\alpha }(\hat{k})\delta u_{\beta }(\hat{l})\delta
t_{2}(-\hat{p})},  \label{a143}
\end{equation}
with a reduced form given by 
\begin{equation}
\Theta _{\alpha \beta }^{(2)}(\hat{k},\hat{l},\hat{p})=(2\pi )^{4}\delta (%
\hat{k}+\hat{l}+\hat{p})P_{\alpha 1}({\bf k})P_{\beta 1}({\bf l})\Theta
^{(2)}(\hat{k},\hat{l}),  \label{a144}
\end{equation}
A standard calculation shows that it is related to $G_{\alpha \beta }^{(2)}$
by 
\begin{equation}
\Theta _{\alpha \beta }^{(2)}(\hat{k},\hat{l},\hat{p})=-\int \Gamma
_{\lambda \alpha }(\hat{q},\hat{k})\,\Gamma _{\mu \beta }(\hat{q}^{\prime },%
\hat{l})\,G_{\lambda \mu }^{(2)}(\hat{q},\hat{q}^{\prime },\hat{p})\,d\hat{q}%
d\hat{q}^{\prime },  \label{a145}
\end{equation}
from which, on making use of (\ref{a141})-(\ref{a144}), we obtain 
\begin{equation}
\Theta ^{(2)}(\hat{k},\hat{l})=-\Gamma (\hat{k})\Gamma (\hat{l})G_{2}(\hat{k}%
,\hat{l}).  \label{a146}
\end{equation}

Next, we average $G_{\alpha \beta }^{(2)}$ over the forcing separation to
get 
\begin{equation}
\overline{G}_{\alpha \beta }(\hat{x})=2\int G_{\alpha \beta }^{(2)}(\hat{k},%
\hat{k})\exp (2i\hat{k}\cdot \hat{x})D\hat{k}.  \label{a147}
\end{equation}
This integral shows that the Fourier transform of $\overline{G}_{\alpha
\beta }^{(2)}(\hat{x})$ depends only on the diagonal components of the
reduced function (\ref{a142}). It follows, therefore, from (\ref{a146}) and (%
\ref{a147}), that the 1PI object which we need to consider, in order to
determine $Z_{2}$, is 
\[
\Theta ^{(2)}(\hat{k},\hat{k})=-\Gamma (\hat{k})^{2}G_{2}(\hat{k},\hat{k}). 
\]
Indeed, an application of (\ref{a42}), together with (\ref{a95}), shows that
its bare and renormalised forms are connected by 
\[
\Theta _{R}^{(2)}(\hat{k},\hat{k})=Z_{2}\Theta _{B}^{(2)}(\hat{k},\hat{k}). 
\]
In this way, as we have indicated, we arrive at a function which can be
renormalised using the $Z_{2}$ counterterm alone.

The normalisation condition for $\Theta ^{(2)}(\hat{k},\hat{k})$ is again
applied at the point $\hat{k}=\hat{m},$ and chosen to be consistent with the
tree level approximation, which gives 
\begin{equation}
\Theta ^{(2)}(\hat{m},\hat{m})=\Theta _{0}^{(2)}(\hat{m},\hat{m})=-1,
\label{a150}
\end{equation}
so that the 1 and 2-loop terms satisfy the normalisation conditions 
\begin{equation}
\Theta _{1}^{(2)}(\hat{m},\hat{m})=\Theta _{2}^{(2)}(\hat{m},\hat{m})=0.
\label{a151}
\end{equation}

The diagrams giving $\Theta ^{(2)}(\hat{m},\hat{m})$ to 2-loop order are
shown in Fig.4. Their new feature is the appearance of the heavy dot vertex.
This represents the $O_{2}$ composite operator vertex, which is shown in
Fig.1({\it iv}) for the general case of $O_{s}$.We can understand how these
diagrams arise from the loop expansion of $K$ by using the general procedure
described in [15]. This depends on the fact that (\ref{a140}) is a special
case of the $4th$ order correlation function of elementary fields defined by 
\[
B_{\alpha \beta \gamma \delta }^{(4)}(\hat{x}^{\prime },\hat{x}^{\prime
\prime },\hat{x},\hat{z})=\frac{i^{2}}{2}\left\langle \widetilde{v}_{\alpha
}(\hat{x}^{\prime })\widetilde{v}_{\beta }(\hat{x}^{\prime \prime
})v_{\gamma }(\hat{x})v_{\delta }(\hat{z})\right\rangle , 
\]
in which the arguments $\hat{x}$ and $\hat{z}$ coalesce. Hence, their
Fourier transforms are related. In particular, the connection between their
respective 1PI functions is 
\[
\Theta _{\alpha \beta }^{(2)}(\hat{k},\hat{l},\hat{m})=\frac{1}{2}\int \Phi
_{\alpha \beta \lambda \mu }(\hat{k},\hat{l},\hat{m}-\hat{q},\hat{q}%
)G_{\lambda 1}(\hat{m}-\hat{q})G_{\mu 1}(\hat{q})\,D\hat{q}, 
\]
where $\Phi $ is the 1PI form corresponding to $B^{(4)}$, which is generated
by 
\[
\Phi _{\alpha \beta \gamma \delta }(\hat{k},\hat{l},\hat{p},\hat{q})=(2\pi
)^{16}\frac{\delta ^{4}K}{\delta u_{\alpha }(\hat{k})\delta u_{\beta }(\hat{l%
})i\delta \widetilde{u}_{\gamma }(\hat{p})i\delta \widetilde{u}_{\delta }(%
\hat{q})}. 
\]
This implies that the diagrams for $\Theta ^{(2)}(\hat{m},\hat{m})$ are
constructed from the diagrams for $\Phi $ by tying the two dotted external
legs of the latter to form the $O_{2}$ vertex.

The 1-loop diagram for $\Theta ^{(2)}(\hat{m},\hat{m})$ shown in Fig.4({\it %
iv}) is constructed from the tree level diagram for $\Phi $, which is shown
opposite to it in Fig.4({\it i}). Similarly, the two 2-loop diagrams for $%
\Theta ^{(2)}(\hat{m},\hat{m})$ , shown in Figs.4({\it v}) and ({\it vi}),
are constructed from the 1-loop diagrams for $\Phi $, again shown opposite
to them in Figs.4({\it ii}) and ({\it iii}). The other possible 2-loop
diagrams for $\Theta ^{(2)}(\hat{m},\hat{m})$ , which arise from the two
remaining 1-loop diagrams for $\Phi $, are discarded because the logarithmic
divergences disappear, after integration over the solid angles. In addition,
diagrams which produce longitudinal terms obviously make no contribution to $%
Z_{2}$ and can also be discarded.

The 1-loop diagram of Fig.4({\it iv}) contributes to $\Theta _{1}^{(2)}$ the
term 
\[
P_{\alpha 1}({\bf m})P_{\beta 1}({\bf m})\Theta _{1}^{(2)\prime }(\hat{m},%
\hat{m})=\int D\hat{p}P_{\lambda \gamma \alpha }({\bf m}-{\bf p})P_{\nu
\delta \beta }({\bf m}+{\bf p})G_{\lambda 1}(-\hat{m}+\hat{p})G_{\nu 1}(-%
\hat{m}-\hat{p})Q_{\gamma \delta }(\hat{p}), 
\]
which yields a logarithmic divergence 
\begin{equation}
\Theta _{1}^{(2)\prime }(\hat{m},\hat{m})=-g\log \left( \frac{\mu }{\kappa }%
\right) .  \label{a156}
\end{equation}
The contractions implied in (\ref{a145}) again permit us to discard the
longitudinal part of the above integral.The counterterm vertex shown in
Fig.2({\it iii}) adds a contribution 
\[
-P_{\alpha 1}({\bf m})P_{\beta 1}({\bf m})\Theta _{1}^{(2)\prime \prime }(%
\hat{m},\hat{m})=-P_{\alpha 1}\left( {\bf m}\right) P_{\beta 1}\left( {\bf m}%
\right) \Delta Z_{2}, 
\]
so that by (\ref{a139}) its contribution to $\Theta _{1}^{(2)}$ is 
\begin{equation}
\Theta _{1}^{(2)\prime \prime }(\hat{m},\hat{m})=-a_{1}^{(2)}g\log \left( 
\frac{\mu }{\kappa }\right) .  \label{a158}
\end{equation}
But, from the normalisation condition (\ref{a151}), we have 
\[
\Theta _{1}^{(2)}(\hat{m},\hat{m})=\Theta _{1}^{(2)\prime }(\hat{m},\hat{m}%
)+\Theta _{1}^{(2)\prime \prime }(\hat{m},\hat{m})=0, 
\]
which, upon substituting (\ref{a156}) and (\ref{a158}), gives 
\begin{equation}
a_{1}^{(2)}=-1.  \label{a160}
\end{equation}

At 2-loop order the diagrams in Figs.4({\it v}) and ({\it vi}) contribute
the terms 
\[
\Theta _{2}^{(2)\prime }(\hat{m},\hat{m})=\frac{9}{4}g^{2}I_{3}, 
\]
and 
\[
\Theta _{2}^{(2)\prime \prime }(\hat{m},\hat{m})=\frac{9}{4}g^{2}I_{4}, 
\]
where 
\[
I_{3}=-\int\limits_{\varepsilon }^{\infty }\int\limits_{\varepsilon
}^{\infty }\frac{2+x+y}{x^{2}y^{2}(1+x)(2+y)(1+x+y)}dxdy, 
\]
and 
\[
I_{4}=-\int\limits_{\varepsilon }^{\infty }\int\limits_{\varepsilon
}^{\infty }\frac{(2+x)(2+x+y)(1+3y+y^{2})-(1+x)y(2+y)(3+x+y)}{%
x^{2}y^{2}(1+x)(2+x)(1+y)^{2}(2+y)(1+x+y)}dxdy. 
\]
The latter yield logarithmic divergences

\begin{equation}
\Theta _{2}^{(2)\prime }(\hat{m},\hat{m})=\frac{9}{4}g^{2}\left( \frac{5}{3}-%
\frac{1}{6}\log 2\right) \log \left( \frac{\mu }{\kappa }\right) ,
\label{a165}
\end{equation}
and 
\begin{equation}
\Theta _{2}^{(2)\prime \prime }(\hat{m},\hat{m})=\frac{9}{4}g^{2}\left( 
\frac{4}{3}+\frac{1}{3}\log 2\right) \log \left( \frac{\mu }{\kappa }\right)
.  \label{a166}
\end{equation}
To these we must add the 2-loop counterterm corresponding to (\ref{a158}),
namely 
\begin{equation}
\Theta _{2}^{(2)\prime \prime \prime }(\hat{m},\hat{m})=-a_{2}^{(2)}g^{2}%
\log \left( \frac{\mu }{\kappa }\right) .  \label{a167}
\end{equation}
But the normalisation condition (\ref{a151}) gives 
\[
\Theta _{1}^{(2)\prime }(\hat{m},\hat{m})+\Theta _{1}^{(2)\prime \prime }(%
\hat{m},\hat{m})+\Theta _{1}^{(2)\prime \prime \prime }(\hat{m},\hat{m})=0, 
\]
which, after substituting (\ref{a165})-(\ref{a167}), yields 
\begin{equation}
a_{2}^{(2)}=7.0.  \label{a169}
\end{equation}

The foregoing can be generalised to arbitrary $s$. In place of (\ref{a143}),
we now consider the general 1PI response function 
\[
\Theta _{\alpha _{1}...\alpha _{s}}^{(s)}(\hat{k}_{1},...,\hat{k}_{s},\hat{p}%
)=(2\pi )^{4(s+1)}\frac{\delta ^{s+1}K}{\delta u_{\alpha _{1}}(\hat{k}%
_{1})...\delta u_{\alpha _{s}}(\hat{k}_{s})\delta t_{s}(-\hat{p})}, 
\]
with a reduced form defined by 
\[
\Theta _{\alpha _{1}...\alpha _{s}}^{(s)}(\hat{k}_{1},...,\hat{k}_{s},\hat{p}%
)=(2\pi )^{4}\delta (\hat{k}_{1}+...+\hat{k}_{s}+\hat{p})P_{\alpha _{1}1}(%
{\bf k}_{1})...P_{\alpha _{s}1}({\bf k}_{s})\Theta ^{(s)}(\hat{k}_{1},...,%
\hat{k}_{s}). 
\]
Then $Z_{s}$ can be found by eliminating the logarithmic divergences from
the diagonal component $\Theta ^{(s)}(\hat{m},\ldots ,\hat{m})$ as above.
The relevant diagrams are again those shown in Figs.4({\it iv})-({\it vi}),
except that the heavy dot now symbolises the $O_{s}$ vertex of Fig.1({\it iv}%
), so the $s-2$ external legs of $O_{s}$ are not shown explicitly. Each
diagram has a symmetry factor $s(s-1)/2$. As this is the only respect in
which these diagrams differ from those just considered, we have the
relation\ 
\begin{equation}
a_{1,2}^{(s)}=\frac{s(s-1)}{2}a_{1,2}^{(2)}.  \label{a172}
\end{equation}
However, this is an approximate result, because it is not valid for diagrams
containing more than 2-loops. But, as we discuss further below, it suffices
for the calculation of low order exponents. Thus, we have now calculated all
the numerical constants that we require for the evaluation of $\zeta _{n}$.

\section{The Scaling Exponents}

\label{section6}

For $n=2$, we have, from (\ref{a59}), 
\[
\zeta _{2}=\frac{2}{3}+\Delta _{2}, 
\]
where, from (\ref{a55}),(\ref{a60}) and (\ref{a139}). 
\begin{equation}
\Delta _{2}=-g_{\ast }(a_{1}^{(2)}+a_{2}^{(2)}g_{\ast }).  \label{a174}
\end{equation}
Substituting the numerical values calculated above, as given in (\ref{a138}%
),(\ref{a160}) and (\ref{a169}), we get 
\[
\Delta _{2}=\frac{46}{37^{2}}=0.0336, 
\]
which yields 
\[
\zeta _{2}=0.70. 
\]

For $n=3$, we shall verify in Section \ref{section8} that the known exact
result 
\[
\zeta _{3}=1, 
\]
holds.

In the general case, for $n>3$, we have from (\ref{a81}) 
\[
\zeta _{n}=\frac{n}{3}-\tau _{n}. 
\]
For even orders $n=2m$, the anomaly is given by (\ref{a82}), 
\begin{equation}
\tau _{n}=\max_{p}\tau _{np}.  \label{a179}
\end{equation}
But, from (\ref{a55}),(\ref{a80}),(\ref{a139}),(\ref{a172}) and (\ref{a174}%
), we have 
\[
\tau _{np}=\left\{ p(p-1)+(n-p)(n-p-1)-2(m-p)[(2(m-p)-1]\right\} \frac{%
\Delta _{2}}{2}. 
\]
A simple calculation shows that the maximum value of this expression is
attained by the two terms in the series (\ref{a72}) with (a) $p=m$ and (b) $%
p=m-1$; which gives for (\ref{a179})

\begin{equation}
\tau _{n}=m(m-1)\Delta _{2}.  \label{a310}
\end{equation}
For odd orders, $n=2m+1,$ the anomaly is given directly by (\ref{a83}),
which yields 
\[
\tau _{n}=m^{2}\Delta _{2}, 
\]
where we have again used (\ref{a55}),(\ref{a139}),(\ref{a172}) and (\ref
{a174}).

The above results have been used to calculate $\zeta _{n}$ up to $n=10$. The
results are shown in Fig.5, together with the experimental data taken from
[16-20]. It can be seen that the agreement is good up to about $n=7$ and
fair beyond, if we allow for the uncertainties in the experimental data
which begin to arise. In particular, it may be noted that the key values $%
\zeta _{2}=0.70$ and $\zeta _{6}=1.8$ are in good agreement with
experimental data, the respective data sets from [16-20] giving for $\zeta
_{2}$ the values $(0.71,0.70,0.71,0.70,0.71)$ and for $\zeta _{6}$ the
values $(1.78,1.8,1.8,1.71,1.71).$ The divergence of the experimental data
at higher orders reflects the fact that the experimental determination of $%
\zeta _{n}$ is not yet fully satisfactory for the reasons given in [20].
Hence, the good agreement between our calculations at higher values of $n$
with the particular data sets from [16-18] must be treated with caution,
particularly as the expression we have derived above is not applicable at
large orders. This limitation stems from the fact that the mean nonlinear
response function, being an average over the forcing configuration, does not
represent the effect of multiple correlations with sufficient accuracy at
large $n$. In addition, the approximation (\ref{a172}), as we have noted,
only holds up to 2-loop order. Indeed, it is evident from the foregoing that
the overall approximation must fail when $ng_{\ast }\sim 1$. However, this
occurs at roughly $n=20$, which is well above the current limit of reliable
experimental data. Equally, the divergence of our theoretical values at
higher values of $n$ from the other two data sets [19,20] could indicate
that the accuracy of our low order approximation is already beginning to
deteriorate at around $n\sim 10.$

\section{Elimination of Sweeping}

\label{section7}

We now return to the question of the power and power$\times $logarithmic
divergences which, up to this point, we have simply discarded. The fact that
power divergences arise when field-theoretic methods are applied to
turbulence, using an Eulerian approach, was noticed originally in [21].
Their origin was subsequently identified as being due to the kinematic
effect of the sweeping of small eddies by large eddies, having an almost
uniform velocity [22,23]. The remedy was to change from an Eulerian to a
Lagrangian description, but this greatly complicates the subsequent analysis
[24]. However, it has been shown that the elimination of sweeping can be
accomplished more simply by transforming to a frame moving with the local
velocity of the large scale eddies at some chosen reference point, [25,26].
We shall show that a similar approach can be used to eliminate the power and
power$\times $logarithmic divergences within the present framework. In this
way, we shall demonstrate that, although we have started out from an
Eulerian formulation, we ultimately obtain quasi-Lagrangian approximations
for the renormalised functions.

The problem, therefore, is to find a sweeping interaction term, $\Delta
L_{s} $, say, which can be used to eliminate the effect of sweeping
convection. To this end, we introduce a uniform convection ${\bf U}$ into $W$
and average over its probability distribution, which we assume to be a
Gaussian distribution $\varpropto \exp (-U^{2}/2U_{0}^{2}).$This adds to $L$
an additional interaction term given by 
\begin{equation}
\Delta L_{U}=-\frac{U_{0}^{2}}{2}\int {\bf l\cdot m\,}\widetilde{{\bf v}}%
{\bf (}\hat{m})\cdot {\bf v}(-\hat{m})\,\widetilde{{\bf v}}{\bf (}\hat{l}%
)\cdot {\bf v(}-\hat{l})\,D\hat{l}D\hat{m},  \label{a200}
\end{equation}
which represents the effect of a random Galilean transformation of the
velocity field. We have not distinguished between ${\bf v}$ before and after
the transformation for consistency with the earlier expressions, such as (%
\ref{a23}), and bearing in mind that the transformation does not affect the
statistical averages required for the structure functions.

To represent diagrammatically the additional terms which arise in the loop
expansion of $W$ after the inclusion of the sweeping interaction term we
need to introduce a new 4-leg `sweeping' vertex of the type shown in Fig.6(%
{\it i}). The two wavevectors $\hat{l}$ and $\hat{m}$ in (\ref{a200}) enter
this vertex along its continuous legs and leave along the dotted legs. A
pair of legs carrying a particular wavevector must also carry the same
vector index to represent the scalar product. Free wavevectors in a diagram
containing one or more of these sweeping vertices are identified, as
previously, by overall wavenumber conservation, together with conservation
at any NS vertex. Each sweeping vertex then contributes a factor $U_{0}^{2}\,%
{\bf l\cdot m}$, where ${\bf l}$ and ${\bf m}$ are the two wavevectors which
enter the vertex along its two continuous legs. In all other respects the
diagrams are to be interpreted in accordance with the rules given in Section 
\ref{section4}.

Consider now the set of diagrams, containing only NS vertices, which are
associated with a particular Green's function or velocity correlator, ${\cal %
G}$, say. Let $C_{NS}$ denote any such diagram contributing to ${\cal G}$.
We shall show that it is possible to generate all power and power$\times $%
logarithmic divergences of any $C_{NS}$ from a single sweeping interaction
of the form (\ref{a200}). Let $C_{U}$ denote any diagram containing at least
one sweeping vertex. If $C_{U}$ contains no NS vertices at all, then it will
only generate power divergences. But if it contains at least one NS vertex,
it will also generate power$\times $logarithmic divergences.The following
topological argument demonstrates that the power divergences of $C_{NS}$ can
be put into 1-1 correspondence with the $C_{U}$ diagrams relating to ${\cal G%
}$.

Each factor $\tau (\kappa )$ (or, equivalently, $\varepsilon ^{-1}$) in a
power divergence of $C_{NS}$ arises because it is possible for a soft
wavevector ${\bf q}$ to flow through a particular velocity correlator
without flowing through the active legs of the two NS vertices which it
connects, as already discussed in Section \ref{section4}. This situation can
be represented diagrammatically by contracting the correlator into a 4-leg
vertex formed by merging the two NS vertices which it links, whilst leaving
the hard lines in tact. This can be demonstrated as follows. First, the new
vertex must consist of two in-coming full lines which carry hard
wavevectors, ${\bf l}$ and ${\bf m}$ (say), and two outgoing dotted lines
along which they leave.This is because two full legs disappear from the
merged NS vertices and wavevectors leave NS vertices along the dotted leg.
Furthermore, after integrating over the directions of the soft wavevector $%
{\bf q}$, the two merged NS vertices generate, through contraction of the
projectors, a factor proportional to the scalar product of the in-coming
hard lines, ${\bf l\cdot m}$, while the two legs of a pair carrying the same
wavevector acquire the same vector index.The final integration over the
wavenumbers then produces the constant 
\[
\frac{1}{6\pi ^{2}}\int\limits_{\kappa }^{\infty }q^{2}D(q)\tau (q)dq=\frac{3%
}{2}g\tau (\kappa )\nu ^{3}. 
\]
So, such a vertex must, in fact, be of the sweeping convection type (\ref
{a200}), with a coefficient given by 
\begin{equation}
U_{0}^{2}=\frac{3}{2}g\tau (\kappa )\nu ^{3},  \label{a201}
\end{equation}
which relates the {\it rms} velocity of the sweeping eddies to the strength
of the nonlinear interaction.Note that $U_{0}$ is scale dependent, as it
depends on the renormalisation scale $\mu $ through $g$ and $\nu $, and,
hence, it differs according to the fluctuation scale on which the RG
focuses. In physical terms, this reflects the fact that the {\it rms}
velocity of the sweeping eddies depends on the scale selected.

Clearly, if a subset of correlators of $C_{NS},$ each of which carries a
soft wavenumber, is contracted into such vertices, in a manner which allows
hard wavevectors to flow through $C_{NS},$ then the result is a diagram
which is identical to one of the $C_{U}$ diagrams. Moreover, it is clear
that there are always exactly as many ways to contract the correlators in $%
C_{NS}$ as there are different $C_{U\text{ }}$diagrams and that their
symmetry factors must match. This argument demonstrates, therefore, the
important point that the power and power$\times $logarithmic divergences
generated by the NS vertex must arise on account of the background of
kinematic sweeping effects. Moreover, we also see that, in order to
eliminate them, it is only necessary to introduce a sweeping interaction
term into $W$ of opposite sign to the one from which they can be generated,
which, according to (\ref{a200}) and (\ref{a201}) yields the sweeping
interaction term 
\begin{equation}
\Delta L_{s}=\frac{3}{4}\frac{g\nu ^{3}}{\tau (\kappa )}\int {\bf l\cdot m\,}%
\widetilde{{\bf v}}{\bf (}\hat{m}{\bf )\cdot v(}-\hat{m}{\bf )\,}\widetilde{%
{\bf v}}{\bf (}\hat{l}{\bf )\cdot v(}-\hat{l}{\bf )\,}D\hat{l}D\hat{m}.
\label{a202}
\end{equation}
Thus, the sweeping vertex shown in Fig.6({\it i}) is taken to represent the
algebraic factor 
\[
\text{Vertex 6({\it i})}=-\frac{3}{2}g\tau (k)\nu ^{3}\,{\bf l\cdot m}. 
\]
Having inserted (\ref{a202}) into $W$ one is then left with only the pure
logarithmic divergences generated by the NS vertex, which, as we have shown,
can be summed using the RG. This justifies our procedure whereby power\ and
power$\times $logarithmic divergences are discarded when calculating
anomalous exponents.

We now illustrate the cancellation of power and power$\times $logarithmic
divergences in concrete terms by eliminating them to 2-loop order from $%
\Gamma (\hat{k}).$ This will demonstrate how the various symmetry factors
match up. Consider first the 1-loop diagram for $\Gamma _{1}(\hat{k})$
arising from the NS vertex. From our previous result (\ref{a203}), we find
that its power divergence is given, at the normalisation point, by 
\begin{equation}
\Gamma _{1}(\text{diagram 3({\it i})})=\frac{3}{2}g\frac{\tau (\kappa )}{%
\tau (\mu )^{2}}.  \label{a204}
\end{equation}
Here the Feynman rules applied to the sweeping vertex yield the single
diagram of Fig.6({\it ii}), as we anticipate from the fact that the NS
vertices in Fig.3({\it i}) can be merged in only one way. In this case, a
trivial calculation yields 
\[
\Gamma _{1}(\text{diagram 6({\it ii)}})=-\frac{3}{2}g\frac{\tau (\kappa )}{%
\tau (\mu )^{2}}, 
\]
which cancels (\ref{a204}), as required.

Explicit verification that there are no power or power$\times $logarithmic
divergences in $\Gamma (\hat{k})$ at 2-loop order is less trivial. Consider
first diagram ({\it iv}{\bf ) }of Fig.3. The power divergences arising from
this diagram follow from (\ref{a205}) which gives 
\begin{equation}
\Gamma _{2}(\text{diagram 3({\it iv})}=-\frac{9}{4}\frac{g^{2}}{\tau (\mu )}%
\left\{ \frac{1}{\varepsilon ^{2}}+\frac{2}{\varepsilon }+\frac{4}{%
\varepsilon }\log \varepsilon \right\} .  \label{a206}
\end{equation}
For this diagram the corresponding sweeping diagrams are diagrams ({\it i})-(%
{\it iii}) of Fig.7. This follows from the Feynman rules and can be checked
from diagram ({\it iv}) of Fig.3 by first contracting its correlators
individually and then together. By applying the Feynman rules to diagram (%
{\it i}) of Fig.7 we obtain, at the normalisation point, 
\begin{eqnarray*}
P_{\alpha \beta }({\bf m})\Gamma _{2}\left( \text{diagram 7({\it i})}\right)
&=&\frac{3}{2}g\tau (\kappa )\nu ^{3}\int {\bf p\cdot (k-p)}P_{\lambda
\sigma \nu }({\bf m})P_{\tau \rho \beta }({\bf m-p}) \\
&&\times Q_{\rho \nu }(\hat{p})G_{\alpha \lambda }(\hat{m})G_{\sigma \mu }(%
\hat{m}-\hat{p})G_{\mu \tau }(\hat{m}-\hat{p}).
\end{eqnarray*}
We can evaluate this integral using the method described in Section \ref
{section4}. This gives 
\begin{eqnarray*}
\Gamma _{2}(\text{diagram 7({\it i}))} &=&\frac{9}{4}\frac{g^{2}}{\tau (\mu )%
}\frac{1}{\varepsilon }\int\limits_{\varepsilon }^{\infty }\frac{dx}{%
x^{2}(x+1)^{2}} \\
&=&\frac{9}{4}\frac{g^{2}}{\tau (\mu )}\left\{ \frac{1}{\varepsilon ^{2}}+%
\frac{2}{\varepsilon }\log \varepsilon +\frac{1}{\varepsilon }\right\} .
\end{eqnarray*}
Diagram ({\it ii}) of Fig.7 yields the same value 
\[
\Gamma _{2}(\text{diagram 7({\it ii}))}=\Gamma _{2}(\text{diagram 7({\it i}%
)).} 
\]
Finally, evaluation of the diagram ({\it iii}) of Fig 7 is trivial and
yields 
\[
\Gamma _{2}(\text{diagram 7({\it iii}))}=-\frac{9}{4}\frac{g^{2}}{\tau (\mu )%
}\frac{1}{\varepsilon ^{2}}. 
\]
Evidently, the sum of these three diagrams cancels (\ref{a206}) exactly.

Similarly, we can show that the sweeping vertex eliminates the power
divergences arising from the second 2-loop diagram, shown in Fig.3({\it iii}%
). From (\ref{a208}), these are given by 
\begin{equation}
\Gamma _{2}(\text{diagram 3({\it iii})})=-\frac{9}{4}\frac{g^{2}}{\tau (\mu )%
}\left\{ \frac{1}{\varepsilon ^{2}}+\frac{5}{2}\frac{1}{\varepsilon }+\frac{4%
}{\varepsilon }\log \varepsilon \right\} .  \label{a207}
\end{equation}
In this case, the corresponding diagrams generated by the sweeping vertex
are diagrams ({\it iv)-(vi)} of Fig.7 which contribute the terms 
\[
\Gamma _{2}(\text{diagram 7({\it iv)}})=\frac{9}{4}\frac{g^{2}}{\tau (\mu )}%
\left\{ \frac{1}{\varepsilon ^{2}}+\frac{1}{\varepsilon }\log \varepsilon
\right\} , 
\]
\[
\Gamma _{2}(\text{diagram 7({\it v)}})=-\frac{9}{4}\frac{g^{2}}{\tau (\mu )}%
\frac{1}{\varepsilon ^{2}}, 
\]
and 
\[
\Gamma _{2}(\text{diagram 7({\it vi)}})=\frac{9}{4}\frac{g^{2}}{\tau (\mu )}%
\left\{ \frac{1}{\varepsilon ^{2}}+\frac{3}{\varepsilon }\log \varepsilon +%
\frac{5}{2}\frac{1}{\varepsilon }\right\} . 
\]
Again, their sum exactly cancels (\ref{a207}). We have thereby verified to
2-loop order that the sweeping interaction eliminates power divergences from
the linear response function.

\section{The Kolmogorov Approximation}

\label{section8}

The fact that it has been possible to calculate the anomalies successfully
by means of perturbation theory stems, in part, from the incorporation of
the Kolmogorov theory into the zero order approximation. As we have seen,
this has been done by replacing the actual viscous quadratic form in $W$,
arising from the NS equations, by a modified quadratic form, characterised
by an effective random stirring force spectrum $D(k)$ and the effective
timescale $\tau (k)$.We now demonstrate that these two functions can be
deduced self-consistently as part of the calculation and confirm that that
they do have the inertial range forms given in (\ref{a184}) and (\ref{a185}).

To determine these functions, we need two conditions. As in [10], one
condition is supplied by evaluating the energy equation to 1-loop order,
which gives the convergent DIA form, corresponding to the so-called line
renormalisation [24]. In the inertial range, it reduces to the condition
that the energy flux across wavenumbers $\Pi _{E}(k)$ is independent of $k$
and equal to the mean dissipation rate $\epsilon $: 
\[
\Pi _{E}(k)=\epsilon . 
\]
Thus, evaluation of $\Pi _{E}(k)$ to 1-loop order gives the well-known
result [24] 
\[
\Pi _{E}(k)=\int\limits_{k}^{\infty }T(p)dp, 
\]
where 
\begin{eqnarray}
T(p) &=&8\pi ^{2}\int \int_{\Delta }dqdr\frac{p^{3}qr}{\tau (p)^{-1}+\tau
(q)^{-1}+\tau (r)^{-1}}  \nonumber \\
&\times &\left\{ b(p,q,r)Q(r)(Q(q)-Q(p))+b(p,r,q)Q(q)(Q(r)-Q(p))\right\} .
\label{a209}
\end{eqnarray}
Here $\Delta $ indicates integration over the region of the $p,q$ plane in
which $p,q,r$ can form a triangle and 
\[
b(p,q,r)=\frac{\left( p^{2}+q^{2}-r^{2}\right) ^{3}}{8p^{4}q^{2}}+\frac{%
r^{4}-\left( p^{2}-q^{2}\right) ^{2}}{4p^{2}r^{2}}. 
\]

The second condition must be deduced from the linear response function.This
is where difficulties have arisen with this approach in the past, when using
an Eulerian framework, because of the {\it ir} divergences arising from
sweeping. On the other hand, it is known that no divergence problems arise
from sweeping convection in the case of the energy equation [24]. However,
we have just shown how these power divergences can be systematically removed
from the response function (and, indeed, all such functions) by means of a
random Galilean transformation of the velocity field. This leaves the
logarithmic divergences which, as we have seen, are to be eliminated from $%
\Gamma (\hat{k})$ using the $Z_{\nu }$ counterterm. Recall that to fix the
finite part of $\Gamma (\hat{k})$, after this renormalisation, we imposed
the normalisation condition (\ref{a210}) which specifies that its tree level
term should be exact at the normalisation scale $\mu $.Thus, after
eliminating sweeping convection, as described in Section \ref{section7}, and
using the 1-loop normalisation condition (\ref{a210}) to eliminate the
logarithmic divergences, we obtain, at an arbitrary wavevector ${\bf k}$
(with $\omega =0$), the renormalised linear response function 
\begin{eqnarray}
\Gamma ({\bf k},0) &=&\tau (k)^{-1}+\frac{\mu ^{2}\tau (\mu )^{3}-k^{2}\tau
(k)^{3}}{6\pi ^{2}\tau (k)\tau (\mu )}\int\limits_{0}^{\infty }\frac{%
p^{2}\tau (p)^{2}D(p)dp}{\left( \tau (k)+\tau (p)\right) \left( \tau (\mu
)+\tau (p)\right) }  \nonumber \\
&&+\frac{\mu ^{2}\tau (\mu )^{2}-k^{2}\tau (k)^{2}}{6\pi ^{2}}%
\int\limits_{0}^{\infty }\frac{p^{2}\tau (p)D(p)dp}{\left( \tau (k)+\tau
(p)\right) \left( \tau (\mu )+\tau (p)\right) }.  \label{a211}
\end{eqnarray}
It is precisely the condition that this expression should, indeed, yield a
finite renormalised value which provides the required second relation, as we
now explain.

In the inertial range limit, we seek scaling solutions with $\tau (k)\propto
k^{-a}$ and $Q(k)\propto k^{b},$ in which case $D(k)=\tau
(k)^{-1}Q(k)\propto k^{a+b}.$ Now standard dimensional analysis shows that
for (\ref{a209}) to hold in these circumstances, we must have $a+2b=-8,$
[24]. Furthermore, if this scaling solution were to produce a
non-renormalisable divergence in the response function, it would arise in
the second term of (\ref{a211}), since we can assume that $a>0.$ To prevent
this from occurring, the coefficient of the integral must be zero, which
requires 
\[
\frac{\tau (k)}{\tau (\mu )}=\left( \frac{\mu }{k}\right) ^{2/3}, 
\]
giving $a=2/3$, and, hence, $b=-11/3,$ so that $a+b=-3.$Thus, these
relations do, in fact, yield the solution (\ref{a184}) and (\ref{a185}),
which we may conveniently re-write as 
\begin{equation}
\tau (k)^{-1}=\beta \epsilon ^{1/3}k^{2/3}  \label{a212}
\end{equation}
and 
\begin{equation}
D(k)=\frac{\alpha }{2\pi }\epsilon ^{2/3}k^{-3}.  \label{a213}
\end{equation}
Therefore, the energy spectrum function 
\[
E(k)=4\pi k^{2}Q(k)=4\pi k^{2}D(k)\tau (k) 
\]
takes the Kolmogorov inertial range form 
\[
E(k)=\alpha \epsilon ^{2/3}k^{-5/3}, 
\]
and the integral in the third term of (\ref{a211}) is, indeed, finite and
yields 
\[
\Gamma ({\bf k},0)=\tau (k)^{-1}\{1-g\log \left( \frac{k}{\mu }\right) \}. 
\]
For present purposes, explicit evaluation of the two constants is
unnecessary, since they ultimately disappear from the calculation of the
exponents, because they only occur through the coupling constant which, as
we have seen, is eventually evaluated in terms of its fixed point value.

Next, we comment briefly on the effect of allowing for the perturbation
terms (\ref{a182}) which give the difference between the modified quadratic
form and the original viscous form. As in [10], we treat these terms as
being of nominal order $g$. Their effect is, firstly, to re-introduce into $%
\Gamma (\hat{k})$ the viscous timescale $\tau _{\nu }\left( k\right) $ which
was replaced by $\tau _{0}(k).$ Secondly, and more significantly, new
divergences appear. However, it is not difficult to show that the divergent
terms which are independent of $h(k)$ and $\overline{\nu }$ sum exactly to
the amount cancelled by the counterterms, as would be expected. In the
inertial range limit $\overline{\nu }$ $\rightarrow 0,$ this leaves the term
arising from $h(k),$which is given by 
\[
\Delta \Gamma =-\frac{k^{2}}{\tau (k)}\int\limits_{0}^{\infty }\frac{%
p^{2}h(p)dp}{\tau (k)^{-1}+\tau (p)^{-1}}. 
\]
Given that the actual stirring force spectrum function $h(k)$ has remained
arbitrary, subject only to the condition that it yields a finite input power
given by 
\[
4\pi \int\limits_{0}^{\infty }p^{2}h(p)dp=\epsilon , 
\]
it is clear that the above integral for $\Delta \Gamma $ must be finite.

Thus, the role of these perturbation terms is not critical as regards
calculating the anomalous exponents, provided that the the spectrum of the
stirring forces is non-zero only at small $k$, as it should be. However,
what we find is that, although forced at large scales, the above solution
behaves in the inertial range as if the fluid were stirred with a force
spectral function $\propto k^{-3}$. In this context, it is interesting to
note that, in a study of the randomly forced NS equations by a stochastic
force with zero mean and variance $\propto k^{-3}$ [27], evidence of
multiscaling of the structure functions has been found. In particular, the
results obtained for the ratios $\zeta _{n}/\zeta _{2}$ with the $k^{-3}$
spectrum have been shown to agree with the values computed from the NS
equations forced at large scales. This, of course, is exactly what one might
expect from the above approximation.The present results are also consistent
with the numerical calculations in [28], which suggest the scaling $\tau
_{L}(k)\propto k^{-2/3}$, as in (\ref{a212}), for the Lagrangian micro
timescale, as opposed to the scaling $\tau _{E}(k)\propto k^{-1}$ for the
Eulerian micro timescale, evidence for which has also been presented in
[29]. As we have seen, the reason why the Lagrangian timescale applies in
the present calculation is because we have eliminated sweeping by referring
the velocity field to a frame moving with the local velocity of the large
scale eddies which prevail at any chosen scale. This extracts the straining
interactions, which shape the spectrum, from the background of convection,
to yield quasi-Lagrangian approximations.

In a sense, this derivation of the Kolmogorov quadratic form is analogous to
a multiple timescale expansion in nonlinear wave theory, where part of the
nonlinear behaviour is incorporated into the linear approximation, eg via a
slowly changing wave amplitude, the variation of which is then determined
from the nonlinear interaction by requiring the absence of secular terms in
the higher order approximation.Here the requirement is similar in that it
demands the absence of non-renormalisable terms in order to determine the
nonlinear behaviour of the modified quadratic form.

An integral part of the Kolmogorov theory is the exact result that in the
inertial range limit 
\begin{equation}
S_{3}(r)=-\frac{4}{5}\epsilon r,  \label{a307}
\end{equation}
[5]. So we conclude this section by verifying that this result follows from
the present treatment.

Using standard symmetry relations, we can express $S_{3}(r)$ in terms of the
longitudinal component of the equal time triple velocity correlator 
\[
B_{\alpha \beta \gamma }({\bf x})=\left\langle v_{\alpha }(0)v_{\beta
}(0)v_{\gamma }({\bf x})\right\rangle , 
\]
giving 
\begin{equation}
S_{3}(r)=6B_{111}(r,0,0).  \label{a305}
\end{equation}
Now the general form of the Fourier transform of $B_{\alpha \beta \gamma 
\text{ }}$must be 
\[
B_{\alpha \beta \gamma }({\bf k})=iF(k)P_{\gamma \alpha \beta }({\bf k}), 
\]
and so $F(k)$ can be expressed in terms of the transfer spectrum $T(k)$ by 
\[
F(k)=\frac{\pi ^{2}}{k^{4}}T(k), 
\]
while $T(k)$ is given to 1-loop order by (\ref{a209}). Substituting these
results in (\ref{a305}) gives 
\[
S_{3}(r)=12i\pi \int \frac{T(k)}{k^{4}}k_{1}\left( 1-\frac{k_{1}^{2}}{k^{2}}%
\right) \exp (ik_{1}r)D{\bf k}. 
\]

This integral can be expanded in powers of $r$ the lowest order term giving 
\[
S_{3}(r)=-12\pi ^{2}r\int \frac{T(k)}{k^{4}}k_{1}^{2}\left( 1-\frac{k_{1}^{2}%
}{k^{2}}\right) D{\bf k.} 
\]
After integrating over the solid angle, we get 
\[
S_{3}(r)=-\frac{4}{5}\int\limits_{\kappa }^{\infty }T(k)dk. 
\]
This latter integral is, of course, the transport power $\Pi _{E}(\kappa )$,
which is a finite quantity at 1-loop order and equal to the mean dissipation
rate, as indicated in above, and, hence, we recover (\ref{a307}).

The correlation function $B_{111}({\bf x})$ also has an important role in
the derivation of the OPEs required for the structure functions with higher
odd orders, as we shall see shortly.

\section{Derivation of the OPEs}

\label{section9}

We give finally the derivation of the dominant terms of the OPEs which we
have used in Section \ref{section3} to obtain the structure function
expansions. We deal first with the expansions required for the higher order
structure functions with orders $n>3$. These can be obtained using the
technique described in [30]. We defer discussion of the particular case $n=2$
until last, because it requires a different approach for the reasons given
in Section \ref{section3}.

We begin by considering the OPE of the general product, defined in (\ref{a68}%
), as it appears in \ the expansion (\ref{a69}) for $S_{n}(r)$, taking first
the case of even orders $n=2m$, with $p=0,1,\ldots ,m,$ namely 
\[
\Lambda _{n-p,p}(\hat{x},r)=\frac{v_{+}^{n-p}v_{-}^{p}}{p!(n-p)!}, 
\]
where, as previously, $v_{\pm }=v_{1}(x\pm r/2,y,z,t),$ and we have used the
definition (\ref{a9}). Let us consider the effect of inserting $\Lambda
_{n-p,p}$ into a correlation function containing an arbitrary set of
elementary fields $v_{\alpha _{1}}(\hat{x}),\ldots ,v_{\alpha _{l}}(\hat{x}%
_{l}),$ as in (\ref{a45}). Then, following the approach of [30], we can
derive the dominant terms which we have used in Section \ref{section3} by
considering how many of the $v_{+}$ fields can be paired with a $v_{-}$
field to form products of lower order correlation functions.

Consider the case $p=m,$ ie 
\[
\langle v_{\alpha _{1}}(\hat{x}_{1})\ldots v_{\alpha _{l}}(\hat{x}%
_{l})\Lambda _{m,m}(\hat{x},r)\rangle . 
\]
Here each $v_{+}$ can be paired with a $v_{-}$ to yield a product term 
\begin{equation}
\left\langle \left( v_{+}v_{-}\right) ^{m}\right\rangle \left\langle
v_{\alpha _{1}}(\hat{x}_{1})\ldots v_{\alpha _{l}}(\hat{x}_{l})\right\rangle
,  \label{a308}
\end{equation}
which corresponds to the presence of a unit operator term in the OPE, [30].
If, instead, we only select $m-1$ pairs of $v_{+}v_{-}$ products, we obtain
a term of the type 
\[
2\left\langle \left( v_{+}v_{-}\right) ^{m-1}\right\rangle \left\langle
v_{\alpha _{1}}(\hat{x}_{1})\ldots v_{\alpha _{l}}(\hat{x}_{l})\left( \frac{%
v_{+}^{2}}{2}\right) \right\rangle . 
\]
Now, in the limit as $r\rightarrow 0$, $v_{+}^{2}/2$ behaves like an
insertion of $O_{2}(\hat{x})$ into the correlation function of elementary
fields [30]. Hence, this product tends to 
\begin{equation}
2\left\langle \left( v_{+}v_{-}\right) ^{m-1}\right\rangle \left\langle
v_{\alpha _{1}}(\hat{x}_{1})\ldots v_{\alpha _{l}}(\hat{x}_{l})O_{2}(\hat{x}%
)\right\rangle .  \label{a309}
\end{equation}
But the averages of powers of $v_{+}v_{-\text{ }}$ simply yield
non-stochastic functions of $r$, which we shall denote generically by $%
C_{0}(r),C_{2}(r),\ldots ,$ as appropriate. Thus, from (\ref{a308}) and (\ref
{a309}), we obtain, in the limit as $r\rightarrow 0$, 
\[
\left\langle v_{\alpha _{1}}(\hat{x}_{1})\ldots v_{\alpha _{l}}(\hat{x}%
_{l})\Lambda _{m,m}(\hat{x},r)\right\rangle =\left\langle v_{\alpha _{1}}(%
\hat{x}_{1})\ldots v_{\alpha _{l}}(\hat{x}_{l})\left[ C_{0}(r)+C_{2}(r)O_{2}(%
\hat{x})+\ldots \right] \right\rangle . 
\]
Since the elementary fields are arbitrary, it follows that we have an OPE of
the form 
\[
\Lambda _{m,m}(\hat{x},r)=C_{0}(r)I+C_{2}(r)O_{2}(\hat{x})+\ldots . 
\]
The point about expansions of this type is that the operators of increasing
complexity do, indeed, produce subdominant terms in the expansion of $%
S_{n}(r).$ Here, for example, the unit operator term, as we have shown,
produces the dominant scaling with anomalous exponent given by (\ref{a310}),
whereas the quadratic term can be readily shown to give the smaller exponent 
$\tau _{n}=[m(m-1)-1]\Delta _{2},$ and, hence, is subdominant, while further
terms in the expansion would produce even greater reductions.

A similar argument applies when $p=m-1.$ In this case, however, we cannot
pair every $v_{+}$ with a $v_{-}.$ Therefore, the unit operator term cannot
appear in the OPE for $\Lambda _{m+1,m-1.}$ If, however, we pair every $%
v_{-} $ with a $v_{+}$ then the remaining $v_{+}^{2}$ pairs with the
elementary fields and, in the limit as $r\rightarrow 0$, again appears as an 
$O_{2}(\hat{x})$ insertion. In this case, therefore, the OPE starts with $%
O_{2}(\hat{x})$ to give 
\[
\Lambda _{m+1,m-1}(\hat{x},r)=C_{2}(r)O_{2}(\hat{x})+\ldots . 
\]
By continuing with this argument, we see that the dominant term of the OPE
for the general case of $\Lambda _{n-p,p}$ must take the form given in (\ref
{a70}).

Consider next odd orders, $n=2m+1$. When $p=m$, we have a term of the form 
\[
\left\langle v_{+}^{2}v_{-}\right\rangle \left\langle \left(
v_{+}v_{-}\right) ^{m}\right\rangle \left\langle v_{\alpha _{1}}(\hat{x}%
_{1})\ldots v_{\alpha _{l}}\left( \hat{x}_{l}\right) \right\rangle , 
\]
which, again, corresponds to the presence of a unit operator term, which is,
thus, the dominant term of the OPE, giving 
\[
\Lambda _{m+1,m}(\hat{x},r)=C_{0}(r)I+\ldots . 
\]
When $p=m-1,$ by pairing each $v_{-}$ with a $v_{+},$ we obtain a term of
the form 
\[
\left\langle \left( v_{+}v_{-}\right) ^{m-1}\right\rangle \left\langle
v_{\alpha _{1}}\left( \hat{x}_{1}\right) \ldots v_{\alpha _{l}}\left( \hat{x}%
_{l}\right) v_{+}^{3}\right\rangle . 
\]
In the limit as $r\rightarrow 0$, $v_{+}^{3}$ appears as an insertion of the
cubic operator $O_{3}(\hat{x}),$ so that here the OPE takes the form 
\[
\Lambda _{m+2,m-1}(\hat{x},r)=C_{3}(r)O_{3}(\hat{x})+\ldots . 
\]
Continuing this process, we get for the next OPE 
\[
\Lambda _{m+3,m-2}(\hat{x},r)=C_{5}(r)O_{5}(\hat{x})+\ldots , 
\]
and so on. But, in fact, the only term which contributes to $S_{n}(r)$ for
odd $n$ is the unit operator term of $\Lambda _{m+1,m}$ because $\
\left\langle O_{2s+1}(\hat{x})\right\rangle =0$ for any integer $s$, in the
case of homogeneous isotropic turbulence.

In the particular case of $v_{+}v_{-}$, we can establish the form of its OPE
by using an expansion in the Fourier domain, in which the wavenumber $q$,
corresponding to the separation $r$, tends to infinity, as described for
instance, in [7,8]. To this end, we start by considering the general
correlation function 
\[
H_{\alpha \beta \lambda \mu }(\hat{x}_{1},\hat{x}_{2}\mid \hat{x}^{\prime },%
\hat{x}^{\prime \prime })=\left\langle v_{\alpha }\left( \hat{x}_{1}\right)
v_{\beta }\left( \hat{x}_{2}\right) v_{\lambda }\left( \hat{x}^{\prime
}\right) v_{\mu }\left( \hat{x}^{\prime \prime }\right) \right\rangle , 
\]
for the case in which $\hat{x}^{\prime }$ and $\hat{x}^{\prime \prime }$
tend to a common point $\hat{x},$ well separated from $\hat{x}_{1}$ and $%
\hat{x}_{2}$. For simplicity of presentation here, we have included only two
arbitrary fields $v_{\alpha }(\hat{x}_{1})$ and $v_{\beta }(\hat{x}_{2}).$%
Denote its Fourier transform by 
\begin{eqnarray}
H_{\alpha \beta \lambda \mu }(\hat{p},\hat{p}^{\prime }\mid \hat{k},\hat{k}%
^{\prime }) &=&\left\langle v_{\alpha }\left( \hat{p}\right) v_{\beta
}\left( \hat{p}^{\prime }\right) v_{\lambda }(\hat{k})v_{\mu }(\hat{k}%
^{\prime })\right\rangle  \label{a312} \\
&=&\left( 2\pi \right) ^{4}\delta (\hat{p}+\hat{p}^{\prime }+\hat{k}+\hat{k}%
^{\prime })\tilde{H}_{\alpha \beta \lambda \mu }(\hat{p},\hat{p}^{\prime
}\mid \hat{k},\hat{k}^{\prime }).  \nonumber
\end{eqnarray}
Then, in terms of the reduced correlation function, we can write 
\begin{eqnarray}
H_{\alpha \beta \lambda \mu }(\hat{x}_{1},\hat{x}_{2} &\mid &\hat{x}^{\prime
},\hat{x}^{\prime \prime })=\int D\hat{p}D\hat{p}^{\prime }D\hat{q}\,\tilde{H%
}_{\alpha \beta \lambda \mu }(\hat{p},\hat{p}^{\prime }\mid \hat{q}-\frac{%
\hat{p}+\hat{p}^{\prime }}{2},-\hat{q}-\frac{\hat{p}+\hat{p}^{\prime }}{2}) 
\nonumber \\
&&\times \exp \left\{ i\hat{p}\cdot \left( \hat{x}_{1}-\frac{\hat{x}^{\prime
}+\hat{x}^{\prime \prime }}{2}\right) +i\hat{p}^{\prime }\cdot \left( \hat{x}%
_{2}-\frac{\hat{x}^{\prime }+\hat{x}^{\prime \prime }}{2}\right) +i\hat{q}%
\cdot \left( \hat{x}^{\prime }-\hat{x}^{\prime \prime }\right) \right\} .
\label{a311}
\end{eqnarray}

When the arguments in (\ref{a312}) coalesce to the common point $\hat{x}$,
we obtain the correlation function 
\[
Q_{\alpha \beta \lambda \mu }(\hat{x}_{1},\hat{x}_{2}\mid \hat{x}%
)=\left\langle v_{\alpha }\left( \hat{x}_{1}\right) v_{\beta }\left( \hat{x}%
_{2}\right) v_{\lambda }\left( \hat{x}\right) v_{\mu }\left( \hat{x}\right)
\right\rangle , 
\]
with Fourier transform 
\begin{eqnarray*}
Q_{\alpha \beta \lambda \mu }(\hat{p},\hat{p}^{\prime } &\mid &\hat{q}%
)=\left\langle v_{\alpha }\left( \hat{p}\right) v_{\beta }\left( \hat{p}%
^{\prime }\right) (v_{\lambda }v_{\mu })\left( \hat{q}\right) \right\rangle
\\
&=&\left( 2\pi \right) ^{4}\delta (\hat{p}+\hat{p}^{\prime }+\hat{q})\tilde{Q%
}_{\alpha \beta \lambda \mu }(\hat{p},\hat{p}^{\prime }\mid \hat{q}).
\end{eqnarray*}
Thus, corresponding to (\ref{a311}), we have 
\[
Q_{\alpha \beta \lambda \mu }(\hat{x}_{1},\hat{x}_{2}\mid \hat{x})=\int D%
\hat{p}D\hat{p}^{\prime }Q_{\alpha \beta \lambda \mu }(\hat{p},\hat{p}%
^{\prime }\mid -\hat{p}-\hat{p}^{\prime })\exp \left\{ i\hat{p}\cdot \left( 
\hat{x}_{1}-\hat{x}\right) +i\hat{p}\cdot \left( \hat{x}_{2}-\hat{x}\right)
\right\} . 
\]

Let $\Psi _{\alpha \beta \lambda \mu }(\hat{p},\hat{p}^{\prime }\mid \hat{k},%
\hat{k}^{\prime })$ and $\Xi _{\alpha \beta \lambda \mu }(\hat{p},\hat{p}%
^{\prime }\mid \hat{q})$ be the 1PI functions associated with the connected
parts of $\tilde{H}_{\alpha \beta \lambda \mu }(\hat{p},\hat{p}^{\prime
}\mid \hat{k},\hat{k}^{\prime })$ and $\tilde{Q}(\hat{p},\hat{p}^{\prime
}\mid \hat{q})$. Denoting the connected part by superscript $c$, we have, as
in Section \ref{section5}, 
\begin{equation}
\tilde{H}_{\alpha \beta \lambda \mu }^{(c)}(\hat{p},\hat{p}^{\prime }\mid 
\hat{k},\hat{k}^{\prime })=-G_{\alpha \alpha ^{\prime }}\left( \hat{p}%
\right) G_{\beta \beta ^{\prime }}\left( \hat{p}^{\prime }\right) G_{\lambda
\lambda ^{\prime }}(\hat{k})G_{\mu \mu ^{\prime }}(\hat{k}^{\prime })\Psi
_{\alpha ^{\prime }\beta ^{\prime }\lambda ^{\prime }\mu ^{\prime }}(\hat{p},%
\hat{p}^{\prime }\mid \hat{k},\hat{k}^{\prime }),  \label{a313}
\end{equation}
and 
\begin{equation}
\tilde{Q}_{\alpha \beta \lambda \mu }^{(c)}(\hat{p},\hat{p}^{\prime }\mid 
\hat{q})=-G_{\alpha \alpha ^{\prime }}(\hat{p})G_{\beta \beta ^{\prime }}(%
\hat{p}^{\prime })\Xi _{\alpha ^{\prime }\beta ^{\prime }\lambda \mu }(\hat{p%
},\hat{p}^{\prime }\mid \hat{q}).  \label{a314}
\end{equation}

According to the standard procedure [7,8], the behaviour of the correlation
function $H_{\alpha \beta \lambda \mu }(\hat{x}_{1},\hat{x}_{2}\mid \hat{x}%
^{\prime },\hat{x}^{\prime \prime })$ as a function of $\hat{x}^{\prime }-%
\hat{x}^{\prime \prime }$, when $\hat{x}^{\prime }$ and $\hat{x}^{\prime
\prime }$ both tend to a common value $\hat{x}$, can be deduced from the
behaviour of $\Psi _{\alpha \beta \lambda \mu }(\hat{p},\hat{p}^{\prime
}\mid \hat{q}-(\hat{p}+\hat{p}^{\prime })/2,-\hat{q}-(\hat{p}+\hat{p}%
^{\prime })/2)$, in the limit as $\hat{q}\rightarrow \infty $, which is,
indeed, apparent from (\ref{a311}) and (\ref{a313}). Now, the diagrams which
contribute to this 1PI correlation function are diagram ({\it i}) of Fig.8,
together with its permutation $(\hat{p},\alpha )\leftrightarrow (\hat{p}%
^{\prime },\beta )$, and diagram ({\it ii}). However, it is easy to see from
these diagrams that, as $\hat{q}\rightarrow \infty $, diagram ({\it ii})
yields a contribution which is smaller than that from diagram ({\it i}) by a
factor $Q_{\sigma \nu }(\hat{q})\sim q^{-11/3}.$ So to derive the dominant
term, we need to focus on diagram ({\it i}) and its permutation. The
corresponding diagrams for $\Xi _{\alpha \beta \lambda \mu }(\hat{p},\hat{p}%
^{\prime }\mid -\hat{p}-\hat{p}^{\prime })$ are diagram ({\it iii}) of Fig.8
plus its permutation $\lambda \leftrightarrow \mu .$

Evaluation of these diagrams using the Feynman rules is straightforward and
yields 
\begin{eqnarray*}
\Psi _{\alpha \beta \lambda \mu }(\hat{p},\hat{p}^{\prime }\mid \hat{q}-%
\frac{\hat{p}+\hat{p}^{\prime }}{2},-\hat{q}-\frac{\hat{p}+\hat{p}^{\prime }%
}{2})= &&\int P_{\lambda \xi \rho }({\bf q-}\frac{{\bf p+p}^{\prime }}{2}%
)P_{\mu \tau \eta }({\bf -q-}\frac{{\bf p+p}^{\prime }}{2})P_{\alpha \gamma
\sigma }({\bf p})P_{\beta \nu \delta }({\bf p}^{\prime }) \\
&&\times Q_{\gamma \delta }(\widehat{s})Q_{\eta \sigma }(\hat{p}-\widehat{s}%
)Q_{\xi \nu }(\hat{p}^{\prime }+\widehat{s})Q_{\rho \tau }(\hat{q}+\frac{%
\hat{p}^{\prime }-\hat{p}}{2}+\widehat{s})D\widehat{s} \\
&&+(\hat{p},\alpha )+(\hat{p}^{\prime },\beta ),
\end{eqnarray*}
and 
\begin{eqnarray*}
\Xi _{\alpha \beta \lambda \mu }(\hat{p},\hat{p}^{\prime }\mid -\hat{p}-\hat{%
p}^{\prime })=-\int P_{\alpha \sigma \gamma }({\bf p}) &&P_{\beta \nu \delta
}({\bf p}^{\prime })Q_{\gamma \delta }(\widehat{s})Q_{\sigma \mu }(\hat{p}-%
\widehat{s})Q_{\lambda \nu }(\hat{p}^{\prime }+\widehat{s})D\widehat{s} \\
+(\lambda \leftrightarrow \mu ). &&
\end{eqnarray*}
Hence, for large $\hat{q},$ we obtain from the last two equations the
relation 
\[
\Psi _{\alpha \beta \lambda \mu }(\hat{p},\hat{p}^{\prime }\mid \hat{q}-%
\frac{\hat{p}+\hat{p}^{\prime }}{2},-\hat{q}-\frac{\hat{p}+\hat{p}^{\prime }%
}{2})=P_{\lambda \xi \rho }({\bf q})P_{\mu \tau \eta }({\bf q})Q_{\rho \tau
}(\hat{q})\Xi _{\alpha \beta \xi \eta }(\hat{p},\hat{p}^{\prime }\mid -\hat{p%
}-\hat{p}^{\prime }). 
\]
Combining this with (\ref{a313}) and (\ref{a314}) yields the approximation 
\begin{equation}
H_{\alpha \beta \lambda \mu }^{(c)}(\hat{p},\hat{p}^{\prime }\mid \hat{q}-%
\frac{\hat{p}+\hat{p}^{\prime }}{2},-\hat{q}-\frac{\hat{p}+\hat{p}^{\prime }%
}{2})=C_{\lambda \mu \xi \eta }(\hat{q})Q_{\lambda \mu \xi \eta }^{(c)}(\hat{%
p},\hat{p}^{\prime }\mid -\hat{p}-\hat{p}^{\prime }),  \label{a315}
\end{equation}
where, to this order, 
\begin{equation}
C_{\lambda \mu \xi \eta }(\hat{q})=P_{\lambda \xi \rho }({\bf q})P_{\mu \tau
\eta }({\bf q})\left| G(\hat{q})\right| ^{2}Q_{\rho \tau }({\bf q}).
\label{a316}
\end{equation}

To obtain the required expansion for $v_{+}v_{-}$, we must take the inverse
Fourier transform of (\ref{a315}) for the particular case $\lambda =\mu =1$
with 
\[
\hat{x}^{\prime }=(x+\frac{r}{2},y,z,t)\text{ \ \ and \ \ }\hat{x}^{\prime
\prime }=(x-\frac{r}{2},y,z,t). 
\]
The coefficient $C_{11\xi \eta }(\hat{x}^{\prime }-\hat{x}^{\prime \prime })$
then depends only upon $r$ and, according to (\ref{a316}), it must have the
form 
\[
C_{11\xi \eta }(r)=\int \frac{q_{\xi }q_{\eta }(q_{2}^{2}+q_{3}^{2})}{q^{4}}%
F(q)\exp (-iq_{1}r)D{\bf q}, 
\]
where $F(q)$ is a function only of the wavenumber $q$. It is clear from this
integral that $C_{11\xi \eta }$ must be diagonal in the indices $\xi ,\eta $%
, and have equal transverse components:$C_{1122}=C_{1133}.$

We now define $Q_{\alpha \beta }^{(L)}(\hat{x}_{1},\hat{x}_{2}\mid \hat{x})$
to be the connected correlation function formed from the elementary fields $%
v_{\alpha }(\hat{x}_{1})$ and $v_{\beta }(\hat{x}_{2})$, with the insertion
of the longitudinal energy operator $O_{2}(\hat{x})$, ie it is the
particular case of (\ref{a47}) with $s=2$ and $l=2$. Then 
\[
Q_{\alpha \beta 11}^{(c)}(\hat{x}_{1},\hat{x}_{2}\mid \hat{x})=2Q_{\alpha
\beta }^{(L)}(\hat{x}_{1},\hat{x}_{2}\mid \hat{x}). 
\]
Similarly, we define $Q_{\alpha \beta }^{(T)}(\hat{x}_{1},\hat{x}_{2}\mid 
\hat{x})$ to be the correlation function with $v_{\alpha }(\hat{x}_{1})$ and 
$v_{\beta }(\hat{x}_{2})$, and the insertion of the transverse energy
operator 
\[
O_{2}^{(T)}(\hat{x})=\frac{1}{2}\left( v_{2}^{2}+v_{3}^{2}\right) . 
\]
Thus, we have 
\[
Q_{\alpha \beta 22}^{(c)}(\hat{x}_{1},\hat{x}_{2}\mid \hat{x})+Q_{\alpha
\beta 33}^{(c)}(\hat{x}_{1},\hat{x}_{2}\mid \hat{x})=2Q_{\alpha \beta
}^{(T)}(\hat{x}_{1},\hat{x}_{2}\mid \hat{x}). 
\]
Finally, we define longitudinal and transverse coefficients by writing 
\[
C_{2}(r)=2C_{1111}(r), 
\]
and 
\[
C_{2}(r)=2C_{1122}(r)=2C_{1133}(r). 
\]
Using these definitions, and taking into account the diagonality of $%
C_{11\xi \eta ,}$ enables us to express the inverse Fourier transform of (%
\ref{a315}), for the case $\lambda =\mu =1$, as 
\[
H_{\alpha \beta 11}^{(c)}(\hat{x}_{1},\hat{x}_{2}\mid \hat{x}%
,r)=C_{2}(r)Q_{\alpha \beta }^{(L)}(\hat{x}_{1},\hat{x}_{2}\mid \hat{x}%
)+C_{2}^{\prime }(r)Q_{\alpha \beta }^{(T)}(\hat{x}_{1},\hat{x}_{2}\mid \hat{%
x}), 
\]
which, in the limit as $r\rightarrow 0$, leads to 
\[
\langle v_{\alpha }(\hat{x}_{1})v_{\beta }(\hat{x}_{2})v_{+}v_{-}\rangle
=\left\langle v_{\alpha }(\hat{x}_{1})v_{\beta }(\hat{x}_{2})\left[ \frac{E}{%
3}+C_{2}(r)O_{2}(\hat{x})+C_{2}(r)O_{2}^{(T)}(\hat{x})+\ldots \right]
\right\rangle . 
\]
Since the fields $v_{\alpha }(\hat{x}_{1})$ and $v_{\beta }(\hat{x}_{2})$
are arbitrary, we may conclude that 
\[
v_{+}v_{-}=\frac{E}{3}I+C_{2}(r)O_{2}(\hat{x})+\ldots . 
\]
Note that we have discarded the transverse operator because it is
subdominant. This follows immediately from the analysis of Section \ref
{section5}. For example, in the case of $O_{2}^{(T)},$ when we calculate the
corresponding value of the constant $a_{1}^{(2)},$ as defined in (\ref{a139}%
), we get twice the value given in (\ref{a160}) for the longitudinal
operator $O_{2}(\hat{x}),$ because, by isotropy, each of the two transverse
components of $O_{2}^{(T)}(\hat{x})$ contributes an amount equal to the
value obtained for $O_{2}(\hat{x})$ and, hence, the right hand side of (\ref
{a174}) then yields an anomalous exponent of $2\Delta _{2}$, indicating that 
$O_{2}^{(T)}(\hat{x})$ makes a subdominant contribution to $S_{2}(r)$. Thus,
we have shown, to within the order $g^{2}$ of the calculation, that the
dominant term of the OPE for $v_{+}v_{-}$ has the form given in (\ref{a44}).

\section{Summary and Discussion}

The fact that it has been possible to demonstrate multiscaling and calculate
anomalous exponents successfully from the generating functional by means of
perturbation theory, notwithstanding the strong nonlinearity of the NS
equations, is attributable to several factors.These include: (1) the use of
a modified quadratic form, which is derived self-consistently from the NS
nonlinearity;(2) the incorporation in the generating functional of the
composite operators which appear in the definition of the general structure
function;(3) the application of OPEs to derive corrections to the Kolmogorov
exponents in terms of the anomalous dimensions of these operators;(4) the
identification of a class of irreducible Green's functions containing
insertions of these operators, which facilitate the calculation of their
anomalous dimensions;(5) the elimination of sweeping convection effects
using a random Galilean transformation of the velocity field; and, finally,
(6) the deduction of the inertial range scaling using an {\it uv} fixed
point of the RG to achieve the required small wavenumber limit. Let us now
consider how each of these factors contributes to overcoming the obstacles
encountered in previous applications of the RG.

The use of the modified quadratic form is an important element in the
success of our calculation, because it provides an accurate initial
approximation, which yields the Kolmogorov distribution in the inertial
range limit. By contrast, in the early work which employed a field theoretic
RG [31], and in subsequent developments of it [32-34], including equivalent
formulations based on [35], reviewed recently in [36], the zero order
approximation is based solely on on the linear terms of the NS equations, as
in a conventional field theory calculation. Because this is a poor
approximation for turbulence, it does not result in a genuine weak expansion
parameter. For example, in the previous applications of RG techniques based
on an expansion in the force spectrum exponent (ie the $\epsilon $%
-expansion), in which the expansion about $\epsilon =0$ is extrapolated to $%
\epsilon =4,$ the value of the coupling constant is not small, at the {\it %
ir }fixed point which is used. Therefore, the accuracy of the expansion is
uncontrolled. Indeed, according to[37], it may even be uncontrolled when $%
\epsilon \ll 1$, and there are problems in establishing its radius of
convergence and the value of $\epsilon $ at which long range driving becomes
technically irrelevant [38].

However, our expansion is of a different nature. First, we do not use an $%
\epsilon $-expansion. Actually, there is no force power spectrum in our
calculation as such.As we showed, the force spectrum $h(k)$ remains in the
calculation as an arbitrary function, subject only to the requirment that it
yields a finite input power. What the modified quadratic form provides,
however, is an apparent force power spectrum $D(k)$, but its exponent is
fixed by the solution (\ref{a213}), and, thus, cannot be varied. Second, we
do not use an {\it ir} fixed point, because we are interested in taking the
short wavelength limit, for which purpose we require an {\it uv} fixed
point.Together, these differences result in a genuinely small coupling
constant $g$, which is about 1/20 at the fixed point, as shown in Section 
\ref{section4}. Hence, our expansion is inherently more accurate than the $%
\epsilon $-expansion. In fact, given that our calculation is carried out to
2-loop order, its errors are controlled at $g^{3}\sim 10^{-4}.$ Another
significant consequence of using the modified quadratic form is that no
convergence problems are encountered in the {\it uv} region. This, together
with the fact that we do not use an $\epsilon $-expansion or an {\it ir}
fixed point, means that none of the ingredients which cause marginality by
power counting in previous applications of the RG [37], are present in our
approach.

On the other hand, there is a similar problem to be faced in the present
calculation.Any fully renormalised theory of turbulence must contain an
infinite number of renormalised functions because it must be equivalent to
the hierarchy of equations for the cumulants.This equivalence has been
demonstrated recently [39]. In fact, each cumulant will have a
representation as a expansion in terms of irreducible renormalised
functions. Thus, one has an infinite set of vertex functions to contend
with. Now, when any one of these irreducible functions is calculated in
perturbation theory using the modified quadratic form, the overall
logarithmic divergence will remain, after sweeping divergences have been
eliminated. So the problem in the present approach amounts to the
resummation of these logarithms. However, we showed in Section \ref{section5}
that this difficulty could be overcome, in relation to multiscaling, by
identifying the infinite sub-class of functions which yields the desired
information relating to anomalous exponents while being, at the same time,
amenable to resummation using the RG.The irreducible inserted nonlinear
Green's functions defined in Section \ref{section5} satisfy both
requirements. Being fully irreducible they give full $n$-point
correlations.However, as we have seen, to render them tractable, it was
expedient to obtain a mean response to forcing at the centroid of the
excitation points. This averaging thus constitutes a closure approximation.
Although this type of closure approximation permits considerable progress to
be made with the calculation of the exponents, the averaging process limits
its applicability to relatively low orders, $n\lesssim 10$, because the
multiple correlations between the apparent forcing at different space-time
points are not then approximated accurately enough at higher orders. Thus, a
different approximation would be required to obtain the asymptotic scaling
at large orders and it remains for future work to discover a suitable
approach.

\bigskip

\bigskip

\ \ \ \ \ \ \ \ \ \ \ \ \ \ \ \ \ \ \ \ \ \ \ \ \ \ \ \ \ \ \ \ \ \ \ \ \ \
\ \ \ \ \ \ \ REFERENCES

\begin{enumerate}
\item  M.Nelkin,``Universality and scaling in fully developed turbulence,''
Adv. Phys. {\bf 43}, 143 (1994).

\item  K.R.Sreenivasan and R.A.Antonia,``The phenomenology of small scale
turbulence,'' Annu.Rev.Fluid Mech. {\bf 29, }435 (1997).

\item  K.R.Sreenivasan,``Fluid Turbulence,'' Rev. Mod. Phys.{\bf 71, }S383
(1999).

\item  A.N.Kolmogorov,``Local structure of turbulence in an incompressible
fluid for very large Reynolds numbers,'' Dokl.Acad. Nauk. SSSR, {\bf 30},
299 (1941).

\item  U.Frisch, {\it Turbulence,} Cambridge University Press (1995).

\item  W.D.McComb, {\it The Physics of Fluid Turbulence, }Clarendon Press,
Oxford (1996).

\item  J.Zinn-Justin, {\it Quantum Field Theory, }Clarendon Press, Oxford
(1996)

\item  J.C.Collins, {\it Renormalisation,} Cambridge University Press (1984).

\item  P.C.Martin, E.D. Siggia and H.A.Rose, ``Statistical dynamics of
classical systems,'' Phys. Rev. A {\bf 8}, 423 (1973).

\item  S.F.Edwards and W.D.McComb, ``Statistical mechanics far from
equilibrium,''J.Phys.A {\bf 2}, 157 (1969).

\item  S.V.Bazdenkov and N.N.Kukharkin, ``On the variational method of
closure in the theory of turbulence,''Phys.Fluids A {\bf 5}, 2248 (1993).

\item  S. Pokorski, {\it Gauge Field Theories}, Cambridge University Press
(1987).

\item  E.V.Teodorovich, ``Diagram equations of the theory of fully developed
turbulence,'' Theor. and Math. Phys. {\bf 101}, 1177 (1994).

\item  M. Le Bellac, {\it Quantum and Statistical Field Theory}, Clarendon
Press, Oxford (1991).

\item  J.J. Binney, N.J. Dowrick, A.J. Fisher, and M.E.J. Newman, {\it The
Theory of Critical Phenomena}, Clarendon Press, Oxford (1992).

\item  R.Benzi, S. Ciliberto, R. Tripiccione, C. Baudet, F. Massaioli and S.
Succi, ``Extended self-similarity in turbulent flows,''Phys. Rev. E {\bf 48}%
, R29 (1993).

\item  W.van de Water, B. van der Vorst and E. van de Wetering,
``Multiscaling of turbulent structure functions,''Europhys. Lett. {\bf 16},
443 (1991).

\item  F.Anselmet, Y. Gagne, E.J. Hopfinger and R.A. Antonia, ``High-order
velocity structure functions in turbulent sheer flows,''J. Fluid Mech. {\bf %
140}, 63 (1984).

\item  F. Belin, P. Tabeling and H. Willaime, ``Exponents of the structure
functions in a low temperature Helium experiment,''Physica D {\bf 93}, 52
(1996).

\item  K.R.Sreenivasan and B.Dhruva, ``Is there scaling in high Reynolds
number turbulence?'' Prog. Theor. Phys. Suppl.{\bf 130}, 103 (1998).

\item  R.H.Kraichnan, ``The structure of isotropic turbulence at very high
Reynolds numbers,''J.Fluid Mech. {\bf 5, }497 (1959).

\item  B.B.Kadomtsev, {\it Plasma Turbulence }Academic Press, Reading,
Massachusetts (1995).

\item  R.H.Kraichnan, ``Lagrangian history closure approximation for
turbulence,'' Phys.Fluids {\bf 8, }575 (1965).

\item  D.C.Leslie. {\it Developments in the theory of turbulence.}Clarendon
Press, Oxford (1973).

\item  V.Yakhot,``Ultraviolet dynamic renormalisation group,'' Phys.Rev. A 
{\bf 23}, 1486 (1981).

\item  V.S.L'vov,``Scale invariant theory of fully developed
turbulence-Hamiltonian approach,''Physics Reports C {\bf 207, }1 (1991).

\item  A.Sain, Manu and R. Pandit, ``Turbulence and multiscaling in the
randomly forced Navier-Stokes equation,''Phys. Rev. Lett. {\bf 81}, 4377
(1998).

\item  Y.Kaneda,T.Ishihara and K Gotoh,``Taylor expansions in powers of time
of Lagrangian and Eulerian two-point two-time velocity correlations in
turbulence,''Phys.Fluids \ {\bf 11, }2154 (1999).

\item  F.Hayot and C.Jayaprakash,``Dynamic structure factors in models of
turbulence,'' Phys. Rev. E {\bf 57}, R4867 (1998).

\item  J.Cardy. {\it Scaling and Renormalisation in Statistical Physics.}%
Cambridge University Press (1997).

\item  C.DeDominicis and P.C.Martin,``Energy spectra of certain
randomly-stirred fluids,'' Phys.Rev. A {\bf 19}, 419 (1979).

\item  J.K.Bhattacharjee,``Randomly stirred fluids, mode coupling theories
and the turbulent Prandtl number,'' J.Phys. A {\bf 21}, L551 (1988).

\item  E.V.Teodorovich,``On the calculation of the Kolmogorov constant in a
description of turbulence by means of the renormalisation group method,''
Sov.Phys. JETP {\bf 69}, 89 (1989).

\item  D.Ronis,``Field theoretic renormalisation group and turbulence,''
Phys.Rev. A {\bf 36}, 3322 (1987).

\item  V.Yakhot and S.A.Orszag,``Renormalisation group analysis of
turbulence,'' Phys.Rev.Lett. {\bf 57}, 1722 (1986).

\item  L.M.Smith and S.L.Woodruff,``Renormalisation group analysis of
turbulence,'' Ann.Rev.Fluid Mech.{\bf 30}, 275 (1998).

\item  G.L.Eyink,``The renormalisation group method in statistical
hydrodynamics,'' Phys.Fluids {\bf 6}, 3063 (1994).

\item  C.Y.Mou and P.B.Weichman,``Multicomponent turbulence, the spherical
limit, and non-Kolmogorov spectra,'' Phys.Rev. E {\bf 52}, 3738 (1995).

\item  V.S.L'vov and I.Procaccia,``Computing the scaling exponents in fluid
turbulence from first principles: the formal setup,'' Physica A {\bf 257},
165 (1998).
\end{enumerate}

\bigskip \newpage

\ \ \ \ \ \ \ \ \ \ \ \ \ \ \ \ \ \ \ \ \ \ \ \ \ \ \ \ \ \ \ \ \ \ \ \ \ \
\ \ \ \ CAPTIONS TO FIGURES

\begin{enumerate}
\item  FIG.1: Components of the diagrams: (i) velocity correlator; (ii)
linear response function; (iii) Navier-Stokesvertex; (iv) composite operator 
$O_{s}$ vertex.

\item  FIG.2: Counterterm vertices associated with the renormalization of
the elementary fields and the compositeoperators.

\item  FIG.3: The 1PI Feynman diagrams for the linear response evaluated in
Section \ref{section4}.

\item  FIG.4: The 1PI diagrams for the nonlinear response functions
evaluated in Section \ref{section5}.

\item  FIG.5: Comparison of the theoretical expression for $\zeta _{n}$
(full line) with experimental data.

\item  FIG.6: (i) The `sweeping' vertex; (ii) the 1-loop `sweeping' diagram
for the linear response function evaluated in Section \ref{section7}.

\item  FIG.7: The 2-loop `sweeping' diagrams for the linear response
function evaluated in Section \ref{section7}.

\item  FIG.8: The 1-loop diagrams for the correlation functions evaluated in
Section \ref{section9} in connection with the OPEs.
\end{enumerate}

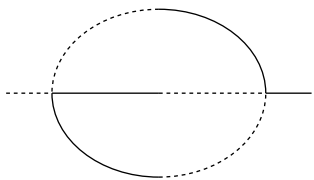
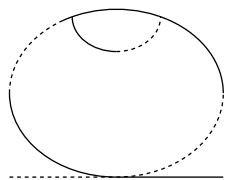
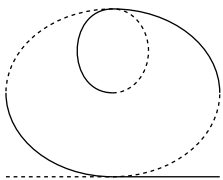
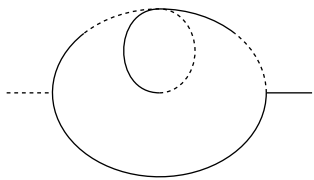
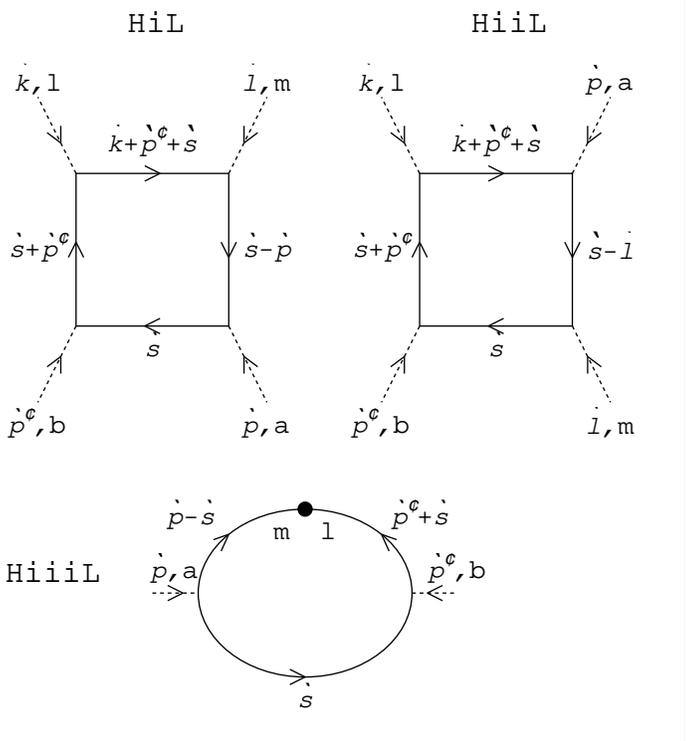
\end{document}